\documentclass[
]{jfmOld}
\usepackage{graphicx}
\usepackage{newtxtext}
\usepackage{newtxmath}
\usepackage{natbib}
\usepackage{hyperref}
\hypersetup{
    colorlinks = true,
    urlcolor   = blue,
    citecolor  = black,
}

\newcommand{\RomanNumeralCaps}[1]
\linenumbers

\usepackage{amsmath,amssymb}
\hypersetup{
	bookmarksnumbered,
	linkcolor=blue,
	anchorcolor = blue,
}
\graphicspath{{./Fig/},{./}}
\usepackage{placeins} 

\usepackage{cleveref}
\crefname{figure}{figure}{figures}
\crefname{equation}{}{}
\Crefname{equation}{Equation}{Equations}
\crefname{section}{\S}{\S\S}

\usepackage{siunitx}
\usepackage[table]{xcolor} 
\definecolor{mycol}{cmyk}{0,1,0,0}  

\title{Transient dispersion in oscillatory flows: auxiliary-time extension method for concentration moments}

\author{
    Weiquan Jiang\aff{1,2}
 \and
    Guoqian Chen \aff{3,1} \corresp{\email{gqchen@pku.edu.cn}}}

\affiliation{
    \aff{1}
    National Observation and Research Station of Coastal Ecological Environments in Macao;
    Macau Environmental Research Institute, Faculty of Innovation Engineering, Macau University of Science and Technology, Macao SAR 999078, China
    \aff{2} Zhuhai MUST Science and Technology Research Institute, Zhuhai 519099, China
    \aff{3} Laboratory of Systems Ecology and Sustainability Science, College of Engineering, Peking University, Beijing 100871, China
}

\begin{document}
\maketitle

\begin{abstract}
The dispersion phenomenon of mass and heat transport in oscillatory flows has wide applications in environmental, physiological and microfluidic flows.
The method of concentration moments is a powerful theoretical tool for analyzing transport characteristics and is well-developed for steady flows.
However, the general solutions of moments derived by Barton (\textit{J.\ Fluid Mech.}, vol.\ 126, 1983, pp.\ 205--218) cannot be applied directly to unsteady flows.
Prior studies needed to re-solve the governing equations of moments from scratch, encountering the complication induced by the time-periodic velocity, leaving higher-order statistics like skewness and kurtosis analytically intractable except for specific cases.
This work proposes a novel approach based on a two-time-variable extension to tackle these challenges.
By introducing an auxiliary time variable, referred to as oscillation time  to characterize the inherent oscillation in the dispersion due to the oscillating flow, the transport problem is extended to a two-time-variable system with a ``steady'' flow term.
This enables the direct use of Barton's expressions and thus avoids the prior complication.
This approach not only offers an intuitive physical perspective for the influence of the velocity oscillation but also clarifies the solution structure of concentration moments.
As a preliminary verification, we examine the transport problem in an oscillatory Couette flow.
The analytical solution agrees well with the numerical result by Brownian dynamics simulations.
The effects of the point-source release and the phase shift of velocity on the transport characteristics are investigated.
By extending the classic steady-flow solution to the time-dependent flows, this work provides a versatile framework for transient dispersion analysis, enhancing predictions in oscillatory transport problems.
\end{abstract}



\section{Introduction}
\label{sec_introduction}

The dispersion phenomenon of mass and heat transport in oscillatory flows has a wide range of applications, from environmental flows in estuaries \citep{holley_dispersion_1970,fischer_mass_1972}, wetlands \citep{zeng_flow_2012} and oceans \citep{yasuda_longitudinal_1984}, to pulsating flows in physiological systems \citep{grotberg_analysis_1990, jiang_bolus_1993,grotberg_biofluid_2004}, to oscillating driving flows in microfluidic systems \citep{vedel_pulsatile_2010,morris_anomalous_2001,ding_shear_2023}, etc.
The time-dependent velocity field introduces additional complexity compared to the case with a steady flow.
Understanding the interplay between diffusion and temporal advection is crucial for predicting and controlling transport processes in various applications.
Taylor dispersion theory \citep{taylor_dispersion_1953} provides a fundamental analytical tool for characterizing the overall transport processes.

The study of the long-time asymptotic dispersion regime for time-dependent flows can date back to the 1960s.
\citet{aris_dispersion_1960} initiated the first analytical investigation of Taylor dispersion in an oscillatory tube flow.
He used the classical method of concentration moments proposed by himself \citep{aris_dispersion_1956} to derive the expression of Taylor coefficient and revealed the fundamental effect of the flow pulsation.
The key to the long-time asymptotic solution is the derivation of the purely periodic part of the first-order moment.
The development of the moment method was promoted by \citet{brenner_macrotransport_1993}, who proposed the generalized Taylor dispersion theory that can be applied to not only time-dependent flows but also to extensions with unsteady components in transport flux.
Alternative methods include using the special solution of concentration distribution to directly derive Taylor dispersivity for oscillatory flows, e.g.\ a linearly longitudinal distribution (although not physically sound for infinitely long channels) \citep{bowden_horizontal_1965, chatwin_longitudinal_1975, watson_diffusion_1983, pedley_effect_1988, mederos_hydrodynamics_2020} and Gaussian distribution for unbounded linear shear flow \citep{young_shear-flow_1982, smith_dispersion_1982}. 
Nevertheless, crucial experimental verification has been undertaken by many researchers \citep{joshi_experimental_1983,jensen_osmotically_2009, song_longitudinal_2015} and more recently by \citet{ding_enhanced_2021}.

Another classical method, the multiscale expansion technique, has also been applied to analyse the long-time asymptotic dispersion coefficient of solute transport in oscillatory flows.
\citet{fife_dispersion_1975} first introduced a two-time-scale general asymptotic solution form of concentration distribution for dispersion problems, which can also treat time-dependent velocity field.
\citet{pavliotis_homogenization_2002} used the typical two-scale homogenization method \citep{pavliotis_multiscale_2008} to derive the Taylor dispersion coefficient for periodic flows.
\citet{ng_time-varying_2004,ng_dispersion_2006} applied the multiple-time-scale homogenization method proposed by \citet{mei_method_1992} to study the cases of  time-periodic channel flows and tube flows with reversible solute adsorption at the walls.
The principles behind these multiscale methods are analogous, although with differences in the specific expansion forms and the selection of scales \citep{camassa_exact_2010,wu_approach_2014,ding_enhanced_2021}.
Recent progress and applications have focused on complex oscillatory transport problems \citep{barik_multi-scale_2019, chu_dispersion_2019, ding_enhanced_2021, karmakar_multi-scale_2023, poddar_effect_2024}.

Recently, the transient dispersion process in time-dependent flows has attracted more interest than the exhaustively explored long-time asymptotic dispersion regime does \citep{vedel_transient_2012}.
The main focus is the complicated temporal evolution of the Taylor--Aris dispersion coefficient.
Thus, the method of concentration moments is probably the remained viable option for analytical studies.
We remark that although the Taylor--Gill generalized dispersion model \citep{gill_dispersion_1971} has also been employed by many studies for various oscillatory flows \citep{gaver_dynamics_1990, hazra_dispersion_1996, rana_solute_2016, wang_concentration_2017,singh_significance_2023},
this method can be considered as a kind of cumulant (alternative to moments) expansion techniques \citep{frankel_generalized_1991,jiang_solution_2018}.
Indeed, the coefficients in Gill's model can be expressed in moments.
Consequently, we will focus on the concentration moment method.

The method of concentration moments is well-developed for transport problems with steady flows.
\citet{barton_method_1983} has derived the general expressions for the full transient solution of moments up to the third order \citep{aquino_equilibrium_2024}, using the method of separation of variables (based on the solution by \citet{aris_dispersion_1956}).
However, for unsteady flows, these results cannot be applied directly.
In fact, the time-dependent velocity makes the solution procedure of moments much more cumbersome.
Because in the ordinary differential equations of the expansion coefficients obtained by separation of variables, the coefficient matrix is time-dependent (i.e.\ the system is non-autonomous) and thus it can considerably increase the complication of the procedure.

Nevertheless, many researchers have aimed to solve the complete  concentration moments in  time-periodic flows.
\citet{yasuda_longitudinal_1982, yasuda_longitudinal_1984} used the method of Green's function to derive the explicit analytical expressions of moments up to the second order for oscillatory currents.
\citet{smith_contraction_1983} obtained the statistics of concentration moments by a Hermite-series representation of concentration distribution.
\citet{mukherjee_dispersion_1988} introduced a so-called time-dependent ``eigenvalue'' problem to extend the approach of \citet{barton_method_1983} to oscillatory flows, using tube and channel flows as examples.
The ``eigenvalue'' problem is in fact in the same form as the cell problem (i.e.\ the long-time asymptotic local transport problem) derived in the homogenization method.
Besides, this form of time-dependent eigenvalue problem was proposed by \citet{shapiro_taylor_1990a} based on Brenner's generalized dispersion theory, despite the lack of intuitiveness.
They pointed out that the eigenvalues could be complex numbers and the eigenfunctions were time-periodic \citep{shapiro_convection_1987}.
Only the zeroth-order moment was discussed for its complete time-dependent solution form, while the higher-order moments were sought in their long-time asymptotic form. 
Actually, if one imposes a new time variable for the oscillatory velocity field, the physical interpretation and the solution structure of the ``eigenvalue'' problem will be fundamentally clear.
This is the primary aim of our work.

More recently, the method of concentration moments has been extended and applied to the study of transient dispersion in complex oscillatory flows.
\citet{vedel_transient_2012} extended earlier studies on single-frequency harmonic pulsatile flows (e.g.\ the work of \citet{mukherjee_dispersion_1988}) to encompass the case of multiple-frequency flows (but with the same base frequency).
A trial expansion form has been constructed to seek the full solution of the first-order concentration moment.
Later, they studied the transient dispersion with a point-source release \citep{vedel_time-dependent_2014}.
For environmental dispersion in tidal wetland flows, \citet{zeng_flow_2012} applied the integral transformation technique to derive the explicit expression for the time-dependent dispersion coefficient, inspiring many subsequent studies \citep{wu_environmental_2012, wang_environmental_2015, guan_transport_2021,das_analysis_2024}.
A later comprehensive study by \citet{ding_enhanced_2021} revisited the transport problem in wall-driven flows (Stokes layer) \citep{bandyopadhyay_unsteady_1999, paul_dispersion_2008}.
They solved the moment equations up to the third order using series expansions.
The analytical results agreed well with their numerical simulations and experimental data.

Apart from the dispersion coefficient, \citet{ding_enhanced_2021} have made the first analytical attempt (to the best of our knowledge) to investigate the temporal evolution of skewness, a fundamental higher-order statistic that reflects the asymmetry of concentration distribution \citep{chatwin_approach_1970,aminian_squaring_2015,aminian_how_2016}.
They also analysed the long-time asymptotic behaviour of skewness: the decay rate for a single-frequency flow is $t^{-3/2}$, instead of $t^{-1/2}$ for the steady case.
This observation is in agreement with the asymptotic analysis by \citet{smith_contaminant_1982, smith_contraction_1983}.

Although the general solution expression of skewness has been provided by \citet{ding_enhanced_2021} for arbitrary time-dependent flows, the derivation was based on a specific initial condition, i.e.\ an instantaneous and uniform source release over the cross-section, such that the zeroth-order concentration moment is $1$ and the solution procedure can be greatly simplified.
Thus, there remains a fundamental need to tackle general initial conditions, particularly for the case with a point-source release.
The corresponding solution can then serve as a Green function \citep{shapiro_taylor_1990a} for arbitrary release scenarios.
\citet{vedel_time-dependent_2014} have investigated the effect of point-source position on the dispersion coefficient.
But its effect on the higher-order statistics including skewness has not been analytically studied for oscillatory flows.

Kurtosis, another useful higher-order statistic to characterize the deviation of concentration distribution from Gaussian distribution, was not addressed by \citet{ding_enhanced_2021}, probably due to the complication introduced by the oscillatory nature of the velocity field.
The long-time asymptotic behaviour of kurtosis has been discussed by \citet{smith_contraction_1983}, concluding that the decay rate for oscillatory flows is $t^{-1}$, the same as that for steady flows \citep{chatwin_cumulants_1972}.
Although for the steady flows, many researchers have analytically investigated the transient evolution of kurtosis \citep{andersson_dispersion_1981,aminian_squaring_2015,wang_basic_2017},
little analytical work has been done for time-periodic flows.
Previous studies had to resort to numerical methods (e.g.\ finite difference method) to calculate the high-order moments for kurtosis \citep{allen_numerical_1982,mazumder_effect_1992,debnath_transport_2019,dhar_dispersion_2022}.

Recently, \citet{singh_significance_2023} and \citet{pal_solute_2025} have derived an integral-form expression of kurtosis.
Their comprehensive and systematic studies have applied the method of concentration moments to investigate the transport problem in pulsating non-Newtonian fluid flows.
They found that the kurtosis of the case with oscillatory flows, similar to skewness, can also be much smaller than that of steady flows, implicating that the concentration distribution may become Gaussian faster for the oscillatory flows than for steady flows \citep{mazumder_effect_1992}.

Although the integral-form expression of kurtosis has been obtained, it should be noted that the solution expression is inexplicit: it contains a lot of unsolved successive integrals with respect to time.
This can be seen in equations (D5)--(D8) in the work of \citet{singh_significance_2023}, where four forms of integrals were defined hierarchically.
Analytically solving these integrals up to the last-order integral is a challenging task.
In fact, they numerically calculated the integration of these integrals, using the Simpson's 1/3 rule.
This numerical solution procedure was previously used by \citet{rana_solute_2016}.
As a comparison, the solution expression of the third-order concentration moment by \citet{ding_enhanced_2021} is explicit with respect to the time variable (i.e.\ no unsolved integrals).
Thus, there remains a need to explicitly solve the fourth-order moment for kurtosis.

This manuscript aims to address the issue of the complication induced by the time-periodic velocity field during the analytical solution procedure of concentration moments, especially when solving for higher-order statistics such as skewness and kurtosis.
We introduce a new auxiliary time variable for the oscillatory velocity field and extend the original transport problem to a two-time-variable problem (which is analogous to the multi-time scale method).
To some extent, the new time dimension can be virtually viewed as an additional ``transverse'' periodic space.
This treatment can not only offer an intuitive physical perspective on the effects of velocity field oscillation on dispersion but also clarify the structure of the governing equations of moments.
More importantly, under this new two-time-variable transport problem, there is no need to re-solve the moment equation.
The general solution expressions obtained by \citet{barton_method_1983} for steady flows can be directly applied to the case of oscillatory flows, because the flow velocity becomes ``steady'' after ``splitting'' the time variable.
The remained procedure simply involves: (i) identifying the new eigenvalue problem and (ii) substituting the eigenvalues and eigenfunctions into Barton's expressions.
Therefore, the complications encountered in previous methods can be avoided.
It is feasible to analytically investigate the temporal evolution of higher-order statistics for time-periodic flows, analogous to the steady-flow case.

\section{Formulation of transport problem and concentration moments}

\subsection{Governing equation}

\begin{figure}
	\centering
	{\includegraphics[width=\textwidth]{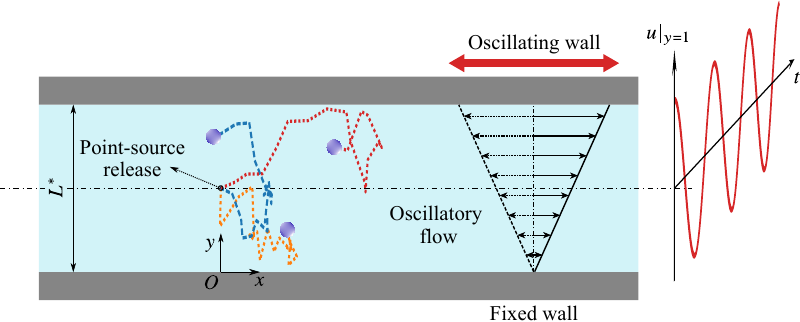}}
	\caption{
		Sketch of solute transport problem in an oscillatory flow created by the oscillating channel wall.
		Several simulated particle trajectories are shown.
		The time variable in the oscillatory velocity field can be virtually viewed as an additional ``transverse'' spatial variable.
		\label{fig_sketch}}
\end{figure}

To illustrate the analytical solution procedure, we consider mass transport in a parallel channel with a time-dependent velocity field.
For simplicity, we focus on the two-dimensional transport problem.
A typical case is shown in \cref{fig_sketch}.

The dimensionless form of the whole transport problem can be adopted as follows:
\begin{subequations} \label{eq_problem_con} 
  \begin{align} \label{eq_concen}
      \frac{\partial C}{\partial t} 
        + u (y, t)  \frac{\partial C}{\partial x} 
      & = \frac{1}{\mathit{Pe}^2} \frac{\partial^2 C}{\partial x^2} 
          + \frac{\partial^2 C}{\partial y^2}, 
  \\ \label{eq_bc_C}
      \left. \frac{\partial C}{\partial y} \right|_{y = 0} &= \left.
      \frac{\partial C}{\partial y} \right|_{y = 1} = 0,
  \\ \label{eq_init_C}
      C | _{t = 0} &= C_{{\text{ini}}} (y) \delta (x) .
  \end{align}
\end{subequations}
The range of the dimensionless variables is  $- \infty < x < + \infty$ (the channel length is infinite), $0 < y < 1$ (the dimensionless channel width is $1$), and $t > 0$ (the time).
Note that the following dimensionless parameters and variables have been chosen (the superscript asterisk denotes dimensional variables and parameters)
\begin{gather}
  x = \frac{1}{\mathit{Pe}} \frac{x^{\ast}}{L^{\ast}}, 
  \quad 
  y =\frac{y^{\ast}}{L^{\ast}},
  \quad 
  t = \frac{t^{\ast}}{{L^{\ast}}^2 /D^{\ast}}, 
\nonumber
\\ \label{eq_dim_var}
  u = \frac{u^{\ast}}{U^{\ast}_{\mathrm{c}}}, 
  \quad 
  C = \mathit{Pe} \frac{C^{\ast}}{C^{\ast}_{{\text{c}}}}, 
  \quad 
  \mathit{Pe} = \frac{U^{\ast}_{\mathrm{c}} L^{\ast}}{D^{\ast}},
  \quad 
  \omega = \frac{\omega^{\ast}}{D^{\ast} / {L^{\ast}}^2}, 
\end{gather} 
where $L^{\ast}$ is the channel width, $D^{\ast}$ is the solute diffusion coefficient, 
$U^{\ast}_{{\text{c}}}$ is a characteristic velocity and $\mathit{Pe}$ is the corresponding Péclet number.
$C^{\ast}_{{\text{c}}}$ is a characteristic concentration constant such that the dimensionless concentration distribution satisfies 
\begin{equation}
  \int^{\infty}_{- \infty} \int^{1}_0 C (x, y, t) \; \mathrm{d} y \mathrm{d} x = 1.
\end{equation}
Namely, $C$ is a well-defined probability density function.

For the advection term in \cref{eq_concen}, the dimensionless time-dependent flow velocity $u (y, t)$ is assumed to be oscillatory and the dimensionless angular frequency is $\omega$.
Namely, the dimensionless period of the flow velocity $T = \frac{2 \pi}{\omega}$, and 
\begin{equation}
  u (y,  t + T) =u (y, t +  \frac{2 \pi}{\omega}) = u (y,  t).
\end{equation}

\subsection{Concentration moments} \label{sec_con_moment}

The goal of this study is to derive the complete time-dependent solution of concentration moments for the above transport problem with unsteady flow velocity.
First, we follow the work by \citet{aris_dispersion_1956} to define the $n$-th order concentration moment for $C (x, y, t)$ as
\begin{equation} \label{eq_def_con_moment}
  C_n (y, t) \triangleq \int^{\infty}_{- \infty} x^n C (x, y, t) \; \mathrm{d} x,
\end{equation}
for $n = 0, 1, 2, \cdots$.
The moments of the cross-sectional mean concentration distribution are more practical in applications.
Here, we use the bar to denote variables related to the cross-sectional mean concentration distribution, and 
\begin{equation}
  \bar{C} (x, t) \triangleq \int^{1}_{0} C (x, y, t) \; \mathrm{d} y.
\end{equation}
Similarly, the $n$-th order moment of $\bar{C}$ is defined as
\begin{equation}
  \bar{C}_n (t) \triangleq \int^{\infty}_{- \infty} x^n \bar{C} (x, t) \; \mathrm{d} x,
\end{equation}
for $n = 0, 1, 2, \cdots$.

The first four order moments are related to the most fundamental overall characteristics of the solute transport problem.
$\bar{C}_0$ represents the total amount of solute mass and $\bar{C}_0=1$ for the current transport problem.
The averaged moving speed of the solute cloud, i.e.\ the drift, is 
\begin{equation}
  u_d \triangleq \frac{\mathrm{d} \overline{\mu_x}}{\mathrm{d} t} 
    =\frac{\mathrm{d}}{\mathrm{d} t} \left( \frac{\bar{C}_1}{\bar{C}_0} \right) 
    = \frac{\mathrm{d} \bar{C}_1}{\mathrm{d} t},
\end{equation}
where $\overline{\mu_x}$ is the mean position of the solute cloud (expected value).

The effective diffusion (apparent dispersivity) is related to the second-order moment $\bar{C}_2$ as
\begin{equation}
  D_e \triangleq \frac{1}{2} \frac{\mathrm{d} \bar{\sigma}}{\mathrm{d} t} =
\frac{1}{2} \frac{\mathrm{d}}{\mathrm{d} t} \left( \frac{\bar{C}_2}{\bar{C}_0}
- \frac{\bar{C}_1^2}{\bar{C}_0^2} \right) = \frac{1}{2}
\frac{\mathrm{d}}{\mathrm{d} t} (\bar{C}_2 - \bar{C}_1^2) .
\end{equation}
where $\bar{\sigma}$ is the variance of the mean concentration distribution.

The higher-order statistics including skewness and kurtosis are also of interest because they can basically reflect the deviation from a Gaussian distribution.
Skewness is a measure of the asymmetry of the longitudinal concentration distribution, whose definition is related to the third-order moment $\bar{C}_3$ as
\begin{equation}
  \mathit{Sk} \triangleq \frac{\bar{\kappa}_3}{\bar{\sigma}^{3}} = \frac{\bar{C}_3- 3 \bar{C}_1 \bar{C}_2 + 2 \bar{C}_1^3}{\bar{\sigma}^{3}},
\end{equation}
where $\bar{\kappa}_3$ is the third-order cumulant.

Kurtosis (excess kurtosis) is a measure of the outliers of a distribution (the tail weight) compared to a Gaussian distribution.
One can use the fourth-order cumulant $\bar{\kappa}_4$ to define the excess kurtosis as
\begin{equation}
  \mathit{Ku} \triangleq \frac{\bar{\kappa}_4}{\bar{\sigma}^{4}} = \frac{\bar{C}_4 - 4 \bar{C}_1 \bar{C}_3 - 3 \bar{C}_2^2 + 12 \bar{C}_1^2 \bar{C}_2 - 6 \bar{C}_1^4}{\bar{\sigma}^{4}}.
\end{equation}
Therefore, we need the fourth-order moment $\bar{C}_4$.

Typically, the next step of the method of concentration moments is to derive the governing equations of moments and then solve them order by order (see the paper by \citet{ding_enhanced_2021} for details).
The governing equation of moments $C_n$ for the current transport problem is given in \cref{sec_eqs_moments}.
However, due to the time-dependent velocity field, the tedious treatment of the time integral with the oscillation terms is a key issue (cf.\ \citet{singh_significance_2023}, with equations (D5)--(D8) numerically integrated).
The corresponding solution procedure is much more complicated than that of the steady-flow case, making it difficult to obtain the complete explicit expressions for higher-order moments.
Therefore, we will not solve the original moment equations.
Instead, we introduce an associated two-time-variable transport problem to get rid of the time-dependent velocity field.

\section{Auxiliary-time extension method}

Here, a new method is proposed to solve the concentration moments.
Analogous to the multi-time scale method, an auxiliary time variable associated with the oscillation frequency of the flow is devised to characterize the intrinsic oscillation in the dispersion.
By extending the original transport problem to a two-time-variable problem, the time-dependent velocity becomes  ``steady'' with the basic time variable while time-dependant only in the auxiliary time dimension.
This key treatment enables us to directly apply the general solution expressions obtained by \citet{barton_method_1983} for steady flows to the case of oscillatory flows.
Generally speaking, it is not necessary to independently re-solve the oscillatory problem.

\subsection{Two-time-variable system}

First, we impose a two-time-variable system with 
\begin{equation} \label{eq_t_0_and_t_1}
  t_0 = t, \quad t_1 = \omega t.
\end{equation}
While $t_0$ represents the original basic time, the new auxiliary time $t_1$, referred to as oscillation time, is introduced directly associated with the frequency $\omega$ of the oscillating flow to characterize the inherent oscillation in the dispersion.
Thus, the oscillatory velocity is only relevant to $t_1$.
For convenience, we define a new velocity function of $t_1$ as
\begin{equation} \label{eq_u_1}
  u_1 (y, t_1) \triangleq u (y, \frac{t_1}{ \omega})=u (y, t),
\end{equation}
such that we can simply replace $u (y, t)$ with $u_1 (y, t_1)$ in the original transport equation \labelcref{eq_concen}.
Note that $u_1$ is a periodic function of $t_1$ with a period of $2 \pi$.

Next, the concentration distribution $C(x,y,t)$ needs to be extended to some virtual ``concentration'' function in the two-time-variable system, denoted as $P(x, y, t_0, t_1)$.
The relationship between $C$ and $P$ is
\begin{equation} \label{eq_C_and_P}
  C (x, y, t) = P (x, y, t_0=t, t_1=\omega t).
\end{equation}
Note that there may exist many possible extensions from $C$ to $P$ in the higher-dimensional system.
One of the most straightforward choices is to set $P$ periodic with respect to $t_1$ because $t_1$ is the oscillation time, and thus it requires
\begin{equation} \label{eq_P_periodic_requirement}
P(x,y,t_0, t_1)= P(x,y,t_0, t_1 + 2\pi).
\end{equation}

For the time derivative, using the chain rule, we have
\begin{equation} \label{eq_t_derivative}
  \frac{\partial C}{\partial t} = \frac{\partial P}{\partial t_0} 
    + \frac{\partial P}{\partial t_1} \frac{\partial t_1}{\partial t} 
    = \frac{\partial P}{\partial t_0} 
    + \omega \frac{\partial P}{\partial t_1} .
\end{equation}

Readers may find that the above two-time-variable procedure is similar to the two-time-scale homogenization method used by previous studies (e.g.\ \citet{ding_enhanced_2021}).
In fact, if one replaces $\omega$ with a so-called small perturbation parameter $\epsilon$, then the formulations are the same.
However, we would not dive into the perturbation analysis with the multi-scale expansion of concentration.
More importantly, the angular frequency of the velocity in this study does not require to be confined as the two-scale method does.
There is no small parameter and no scale analysis.
Thus, we use the term ``two-time-variable'' instead of ``two-time scale''. 
What we are looking for is the complete solution of concentration moments, not the long-time asymptotic concentration distribution (which is the focus of the homogenization method).
Our work finds that the two-time variable formulation could provide a new perspective for the analytical solution of concentration moments.

\subsection{Governing equation for the two-time-variable system}

\Cref{eq_C_and_P} offers the relationship between $C$ and $P$.
Substituting $P$ into the transport equation \labelcref{eq_concen} and using the time derivatives in \cref{eq_t_derivative}, we can obtain the governing equation for $P$ as
\begin{equation} \label{eq_govern_P}
  \frac{\partial P}{\partial t_0} + \omega \frac{\partial P}{\partial t_1} 
    + u_1 (y, t_1)  \frac{\partial P}{\partial x} 
  = \frac{1}{\mathit{Pe}^2} \frac{\partial^2 P}{\partial x^2} 
  + \frac{\partial^2 P}{\partial y^2}.
\end{equation}
Apparently, the second term on the left side reflects the auxiliary flux change due to periotic oscillation in the oscillation time dimension $t_1$, actually a translation effect in the oscillation space, with the oscillation frequency standing for the translation speed in the oscillation time dimension.

Because $P$ is periodic with respect to $t_1$ as in \cref{eq_P_periodic_requirement}, we can confine the whole problem within one periodic interval $[0, 2 \pi]$ of $t_1$.
Namely, $t_1$ is with the modulus $2 \pi$, 
\begin{equation}
  t_1 = (\omega t) \mod 2 \pi,
\end{equation}
then we can impose the following periodic conditions with respect to $t_1$ for $P$
\begin{equation} \label{eq_periodic_P}
  P |_{t_1=0} =  P |_{t_1 = 2 \pi}.
\end{equation}
It is further required that $P$ represents the ``probability density function'' in the two-time-variable system (within one period of $t_1$), namely,
\begin{equation} \label{eq_P_probability_density}
  \int^{\infty}_{- \infty} \int^{1}_0 \int^{2 \pi}_0 P (x, y, t_0, t_1) \; \mathrm{d} t_1 \mathrm{d} y \mathrm{d} x = 1.
\end{equation}
We can define the probability moment of $P$ in \cref{sec_prob_moment}.

The boundary condition of $P$ at the channel walls is the same as that of $C$ in \cref{eq_bc_C}, 
\begin{equation} \label{eq_bc_P}
  \left. \frac{\partial P}{\partial y} \right|_{y = 0} 
  = \left. \frac{\partial P}{\partial y} \right|_{y = 1} = 0.
\end{equation}

To complete the transport problem of $P$ in the two-time-variable system, an ``initial'' condition at $t_0=0$ is needed.
According to the initial condition \labelcref{eq_init_C} of $C$, the simplest choice for $P$ is 
\begin{equation} \label{eq_init_P}
  P | _{t_0 = 0} = C_{{\text{ini}}} (y) \delta (x).
\end{equation}
Namely, the initial condition is independent of $t_1$.
In fact, we just set the initial condition for $P$ to satisfy
\begin{equation}
  C (x, y, t=0) = P (x, y, t_0=0, t_1=0),
\end{equation}
and $P | _{t_0 = 0}$ is periodic with respect to $t_1$.

We have obtained the whole initial--boundary value problem \labelcref{eq_govern_P,eq_periodic_P,eq_bc_P,eq_init_P} of $P$ in the two-time-variable system.
If a solution of $P$ is given, then through \cref{eq_C_and_P}, the solution of $C$ can be immediately obtained because it satisfies all the equations in the transport problem \labelcref{eq_problem_con} of $C$.
However, obtaining the exact solution of $P$ is challenging, as is the case for $C$.
Therefore, we consider the concentration moments instead, as in \cref{sec_con_moment}.

\subsection{Probability moments}
\label{sec_prob_moment}

Similar to the definition of concentration moment in \cref{eq_def_con_moment}, one can also define the moments of $P(x, y, t_0, t_1)$ and then obtain the corresponding moment equations.
The $n$-th order probability moment is 
\begin{equation}
  P_n (y, t_0, t_1) \triangleq \int^{\infty}_{- \infty} x^n P(x, y, t_0, t_1) \; \mathrm{d} x,
\end{equation}
for $n = 0, 1, 2, \ldots$.
According to the relationship \labelcref{eq_C_and_P} between $P$ and $C$, 
we have the relationship between $P_n$ and $C_n$,
\begin{equation} \label{eq_C_and_P_moment}
  C_n (y, t) = P_n(y, t_0=t, t_1=\omega t).
\end{equation}
Namely, once we solve $P_n$, the solution for $C_n$ can be immediately obtained.

According to \cref{eq_govern_P,eq_periodic_P,eq_bc_P,eq_init_P}, the whole initial--boundary value problem for $P_n$ is
\begin{subequations} \label{eq_problem_Pn}
  \begin{align} \label{eq_govern_Pn}
    \frac{\partial P_n}{\partial t_0} + \omega \frac{\partial P_n}{\partial t_1} 
    - \frac{\partial^2 P_n}{\partial y^2} 
    & = n \, u_1 (y, t_1) P_{n - 1} + \frac{n (n-1)}{\mathit{Pe}^2}P_{n - 2},
  \\
    P_n |_{t_1 = 0} &=  P_n |_{t_1 = 2 \pi},
  \\
    \left. \frac{\partial P_n}{\partial y} \right|_{y = 0} 
      &= \left. \frac{\partial P_n}{\partial y} \right|_{y = 1} = 0,
  \\ \label{eq_init_Pn}
    P_n | _{t_0 = 0} & = 
      \left\{\begin{array}{ll}
        C_{{\text{ini}}} (y), & n = 0\\
        0 , & n = 1, 2, \ldots,
      \end{array} \right.
  \end{align}
\end{subequations}
with auxiliary terms $P_{-1}=P_{-2}=0$.

\subsection{Eigenvalue problem}

In the two-time-variable system, the governing equation \labelcref{eq_govern_Pn} of moments for the oscillatory flow is similar to the case of a steady flow because $t_1$ can be viewed as a spatial variable and the velocity is independent of $t_0$.
Therefore, now we are able to directly apply the generic solution expressions of moments given by \citet{barton_method_1983,barton_asymptotic_1984} for the case of steady flows.
The key is to solve the corresponding eigenvalue problem to determine the eigenvalues and eigenfunctions in Barton's expressions.

According to \cref{eq_problem_Pn}, the complete eigenvalue problem for probability moment $P_n$ is
\begin{subequations} \label{eq_eigen_problem}
  \begin{align} 
    - \omega \frac{\partial f}{\partial t_1} +  \frac{\partial^2 f}{\partial y^2} & = - \lambda f,
  \\
    f |_{t_1=0} &=  f |_{t_1 = 2 \pi},
  \\
    \left. \frac{\partial f}{\partial y} \right|_{y = 0} 
      &= \left. \frac{\partial f}{\partial y} \right|_{y = 1} = 0,
  \end{align}
\end{subequations}
where $\lambda$ is an eigenvalue and $f=f(y, t_1)$ is the corresponding eigenfunction.

As mentioned in the introduction, we remark that \citet[equation (3.1)]{mukherjee_dispersion_1988} have introduced a similar ``eigenvalue'' problem (time-dependent) to extend Barton's method to oscillatory flows.
Furthermore, \citet[equations (16) and (17)]{shapiro_taylor_1990a} proposed the same form of the time-dependent problem.
They pointed out that the eigenvalue problem is non-self-adjoint and gave the full solution form of the zeroth-order moment.

The fundamental difference we make here is the introduction of the oscillation time variable $t_1$ to make the oscillatory flow ``steady'' with respect to time variable $t_0$.
The eigenvalue problem \labelcref{eq_eigen_problem} is thus independent of time $t_0$, intrinsically different from the time-dependent one discussed by \citet{mukherjee_dispersion_1988} and \citet{shapiro_taylor_1990a}.
Thus, the introduction of $t_1$ not only enables one to use Barton's result directly, but also can clarify the solution structure of moments (cf.\ \cref{sec_phase_shift}). 

The eigenvalue problem \labelcref{eq_eigen_problem} can be solved using the complex form of Fourier series.
The solution of eigenfunction can be written as
\begin{equation} \label{eq_eigenfunction}
  f_{m, k} (y, t_1) = [a_m \cos (m \pi y) + b_m \sin (m \pi y)] 
  \frac{1}{\sqrt{2 \pi}} \mathrm{e}^{\mathrm{i} k t_1},
\end{equation}
where indexes $k = 0, \pm 1, \pm 2, \ldots$ and $m = 0, 1, 2, \ldots$.
The imaginary unit is $\mathrm{i}$ and $\mathrm{i}^2=-1$.
The factor $\frac{1}{\sqrt{2 \pi}}$ for the Fourier series of $t_1$ and the undermined coefficients $a_m$ and $b_m$ for the function of $y$ are used for the normalization condition \labelcref{eq_othorgonal}.
The corresponding eigenvalue is
\begin{equation} \label{eq_eigenvalue}
  \lambda_{m, k} =  (m \pi)^2 + \mathrm{i} \omega k.
\end{equation}
Namely, the ``eigenvalue'' can be a complex number instead of a real number,
because the eigenvalue problem can be non-self-adjoint.

The inner product for the eigenvalue problem is defined as
\begin{equation}
  \langle f_1, f_2 \rangle 
    \triangleq 
    \int^{2 \pi}_0 \int^1_0 f_1^{\star} (y, t_1) f_2 (y, t_1) \; \mathrm{d} y \mathrm{d} t_1,
\end{equation}
where $f_1$ and $f_2$ are arbitrary functions.
The superscript $\star$ denotes the complex conjugate.
Note that due to the complex form of Fourier series, the orthonormal relationship (orthogonal and normalized) between eigenfunctions can be obtained:
\begin{equation} \label{eq_othorgonal}
  \langle f_{m_1, k_1}, f_{m_2, k_2} \rangle = \delta_{m_1 m_2} \delta_{k_1 k_2} ,
\end{equation}
because
\begin{equation}
  \int^{2 \pi}_0 \left( \frac{1}{\sqrt{2 \pi}} \mathrm{e}^{\mathrm{i} k_1 t_1}
  \right)^{\star} \frac{1}{\sqrt{2 \pi}} \mathrm{e}^{\mathrm{i} k_2 t_1}
  \; \mathrm{d} t_1 = \frac{1}{2 \pi}  \int^{2 \pi}_0
  \mathrm{e}^{\mathrm{i} (- k_1 + k_2) t_1} \; \mathrm{d} t_1 =
  \delta_{0 (- k_1 + k_2)} = \delta_{k_1 k_2} .
\end{equation}
Here, $\delta_{i j}$ is the Kronecker delta,
\begin{equation*}
  \delta_{i j} = \left\{ \begin{array}{ll} 1, & i = j \\ 0, & i \neq j \end{array} \right..
\end{equation*}

\subsection{Solutions of moments}
\label{sec_solution_moments}
The final step of solving $P_n$ is to substitute the eigenvalues \labelcref{eq_eigenvalue} and eigenfunctions \labelcref{eq_eigenfunction} into the generic solution expressions of \citet{barton_method_1983,barton_asymptotic_1984}.
Some recent applications \citep{wang_basic_2017,yang_migration_2021,jiang_analytical_2022} have also given detailed expressions.
Because the expressions are similar (only the notations are different), here, we just discuss $P_0$ and $P_1$ as illustrative examples, while the solutions of the higher-order moments are given in \cref{sec_solution_moments_2_and_higher}.
We mainly follow \S4.4 by \citet{jiang_analytical_2022} and reformulate the solution expressions using the current notation.

First, we can reindex the eigenvalues and eigenfunctions for convenience. 
Note that due to the introduction of $t_1$, there are two indexes $m$ and $k$ for the eigenfunctions ($f_{m, k}(y, t_1)$, with two dimensions).
Hereafter, we use index $i$ for the list of eigenfunctions $\{f_i(y, t_1)\}_{i=0}^{\infty}$ and the list of eigenvalues $\{\lambda_i\}_{i=0}^{\infty}$.
Notice that the index $i$ should not be confused with the imaginary unit $\mathrm{i}$.

For the zeroth-order moment, based on equation (3.1) of \citet{barton_method_1983}, 
\begin{equation} \label{eq_solution_P0}
	P_0(y, t_0, t_1) = \sum^{\infty}_{i = 0} a_i \mathrm{e}^{- \lambda_i t_0} f_i(y, t_1),
\end{equation}
where $a_i$ is the coefficient related to the initial condition,
\begin{equation} \label{eq_a_i}
	a_i \triangleq \langle f_i, P_0 | _{t_0 = 0} \rangle 
  = \int^{2 \pi}_0 \int^1_0 f_i^{\star} (y, t_1) C_{{\text{ini}}} (y) \; \mathrm{d} y \mathrm{d} t_1,
  \quad i = 0, 1, \ldots .
\end{equation}
Note that the initial condition \labelcref{eq_init_Pn} is independent of $t_1$ and thus the coefficient $a_i$ for the oscillatory term with $t_1$ is zero, namely, there is no oscillation in \cref{eq_solution_P0}.
Although we introduce a more complicated function base ($f_i(y, t_1)$), $P_0$ is independent of $t_1$.
Besides, according to \cref{eq_C_and_P_moment}, the oscillatory term in $\mathrm{e}^{- \lambda_i t_0}$ is cancelled out with the term of $t_1$ in $f_i(y, t_1)$, namely, $\mathrm{e}^{- \mathrm{i} \omega k t_0}  \mathrm{e}^{\mathrm{i} k t_1} =1$.
We obtain the same conclusion: no oscillation for $P_0$.
This is not surprising because the zeroth-order moment is the total amount of mass in each parallel streamline (cf.\ the concept of streamwise dispersion \citep{guan_streamwise_2024}), which is not affected by the oscillatory flow.

For the first-order moment $P_1$, based on equation (2.9) by \citet{barton_asymptotic_1984},
\begin{equation} \label{eq_solution_P1}
	P_1(y, t_0, t_1) = 
	t_0 \sum^{\infty}_{i=0} a_i B_{i, i}  \mathrm{e}^{- \lambda_i t_0} f_i(y, t_1)
	+
	\sum^{\infty}_{\substack{i,j=0 \\ j \neq i}}  \frac{a_j B_{i, j} (\mathrm{e}^{- \lambda_j t_0} - \mathrm{e}^{-\lambda_i t_0}) }{\lambda_i - \lambda_j} f_i(y, t_1),
\end{equation}
where $B_{i, j}$ is the expansion coefficient related to the inner product with the velocity profile,
\begin{equation}
  \label{eq_B_ij}
	B_{i, j} \triangleq \langle f_i, u_1 f_j \rangle
  = \int^{2 \pi}_0 \int^1_0 f_i^{\star} (y, t_1) u_1 (y, t_1) f_j (y, t_1) \; \mathrm{d} y \mathrm{d} t_1,
\end{equation}
for $i, j = 0, 1, \ldots$.
Similar to $P_0$, note that the first term in \cref{eq_solution_P1} is independent of $t_1$.
The second term may be oscillatory because it contains $a_j B_{i, j} \mathrm{e}^{- \lambda_j t_0} f_i(y, t_1)$ (when $i \neq j$).

Note that in practice, the oscillatory velocity profile can be generally expressed in the form of the Fourier series of $t_1$ (or $t$ in the original transport problem).
Therefore, the following formula for the integration of the product of three eigenfunctions with respect to $t_1$ can be helpful,
\begin{equation}
  \int^{2 \pi}_0 \frac{1}{\sqrt{2 \pi}} \mathrm{e}^{\mathrm{i} k_1 t_1} 
  \frac{1}{\sqrt{2 \pi}} \mathrm{e}^{\mathrm{i} k_2 t_1}  \frac{1}{\sqrt{2
  \pi}} \mathrm{e}^{\mathrm{i} k_3 t_1} \; \mathrm{d} t_1 = \frac{1}{\sqrt{2
  \pi}} \delta_{0 (k_1 + k_2 + k_3)} .
\end{equation}
Namely, when $k_1 + k_2 + k_3 = 0$, the integration result is
$\frac{1}{\sqrt{2 \pi}}$. 

The solutions for the second- and higher-order moments $P_n$ are given in \cref{sec_solution_moments_2_and_higher}.
Once we solve $P_n$, the solution for $C_n$ can be immediately obtained through \labelcref{eq_C_and_P_moment}.
Then the dispersion characteristics can be calculated according to their definitions in \cref{sec_con_moment}.

\section{Verification with numerical simulation: oscillatory Couette flow}
\label{sec_result}

As a first step of verification and illustration of the new method, the oscillatory Couette flow (Stokes--Couette flow), one of the simplest oscillatory flows, is considered.
To verify the analytical solution of concentration moments, the numerical results by the Brownian dynamics simulation method are used for comparison.

\subsection{Velocity profile of oscillatory Couette flow}

A sketch of the transport problem is shown in \cref{fig_sketch}. 
See also the recent experiment by \citet{ding_enhanced_2021}.
The bottom wall ($y^{\ast}=0$) is at rest while the upper wall ($y^{\ast}=L^{\ast}$) is executing an oscillatory motion with velocity $U_A^{\ast} \cos (\omega^{\ast} t^{\ast})$, where $U_A^{\ast}$ is the amplitude of the velocity oscillation and $\omega^{\ast}$ is the angular frequency.

Based on the dimensionless parameters and variables in \cref{eq_dim_var}, we introduce two additional dimensionless parameters:
\begin{equation}
  \label{eq_dim_var_vel}
  \mathit{Sc} =\frac{\nu^{\ast}}{D^{\ast}}, \quad
  \mathit{Wo} = L^{\ast} \sqrt{\frac{\omega^{\ast}}{\nu^{\ast}}}. 
\end{equation}
Notice that $\omega  = \mathit{Sc} \mathit{Wo}^2$, where $\mathit{Sc}$ is the Schmidt number and $\mathit{Wo}$ is the Womersley number.
$\nu^{\ast}$ is the kinematic viscosity of the fluid.
For the choice of the characteristic velocity in \cref{eq_dim_var}, we use the velocity oscillation scale, namely $U^{\ast}_{c} = U_A^{\ast}$.

The dimensionless Navier--Stokes equation and the boundary conditions for this case are
\begin{equation}
  \frac{1}{\mathit{Sc}} \frac{\partial u}{\partial t} 
  = \frac{\partial^2 u}{\partial y^2},
  \quad
  u |_{y = 1} = \cos (\omega t),
  \quad 
  u |_{y = 0} = 0.
\end{equation}
The oscillatory velocity profile has a solution \citep{ding_enhanced_2021,drazin_navier-stokes_2006}:
\begin{equation}
  \label{eq_velocity_profile}
  u (y, t) = {\Real}\left[\mathrm{e}^{\mathrm{i}
  \omega t} \frac{\sinh \left( \mathrm{e}^{\frac{\pi}{4} \mathrm{i}} 
    \mathit{Wo} y \right)}{\sinh \left( \mathrm{e}^{\frac{\pi}{4}
    \mathrm{i}}  \mathit{Wo} \right)} \right].
  \end{equation}
Notice that $\mathrm{e}^{\frac{\pi}{4} \mathrm{i}} = \sqrt{\mathrm{i}} =\frac{\sqrt{2}}{2} (1+\mathrm{i})$.
$\Real$ means taking the real part of the complex number.

For the two-time-variable system \labelcref{eq_govern_P}, we need to rewrite the velocity profile \labelcref{eq_velocity_profile} in the form of the Fourier series of $t_1$ (see also the definition of $u_1$ in \cref{eq_u_1}) in order to calculate the expansion coefficient $B_{i, j}$ in \labelcref{eq_B_ij}: 
\begin{equation} \label{eq_u_1_oscil_couette}
    u_1 (y, t_1) = \sum_{k = \pm 1}  \frac{\sqrt{2 \pi}}{2}  \left[
  \frac{1}{\sqrt{2 \pi}} \mathrm{e}^{\mathrm{i} k t_1}  \frac{\sinh \left(
  \mathrm{e}^{\frac{\pi}{4} \mathrm{i}} \sqrt{k} \mathit{Wo} y \right)}{\sinh
  \left( \mathrm{e}^{\frac{\pi}{4} \mathrm{i}} \sqrt{k} \mathit{Wo} \right)}
  \right],
  \nonumber
\end{equation}
see also \citet[equation (5)]{ding_enhanced_2021}.
Note that only one frequency is considered here.
For the multiple-frequency case with more Fourier components (see also the paper by \citet{vedel_transient_2012}), the solution procedure is similar but with more expansion coefficients for the ``time'' variable $t_1$.

\begin{table}
	\centering
	\begin{tabular}{cccc}
		Parameter & Description & Reference value & Unit\\
		$L^{\ast}$ & Channel width & $\num{1.6d-3}$ & $\unit{m}$\\
		$D^{\ast}$ & Diffusion coefficient & $\num{8.81d-10}$ & $\unit{m^2s^{-1}}$\\
		$\omega^{\ast}$ & Angular frequency & $\tfrac{2 \pi}{1500}$ & $\unit{rad /s}$\\
		$\nu^{\ast}$ & Kinetic viscosity & $\num{1.13d-6}$ & $\unit{m^2 /s}$\\
		$U_A^{\ast}$ & Amplitude of velocity oscillation & $\num{4.19d-5}$ & $\unit{m /s}$\\
	&  &  & \\
		$\omega$ & Dimensionless angular frequency & $12.17$ & \\
		$\mathit{Pe}$ & Péclet number & $76.07$ & \\
		$\mathit{Wo}$ & Womersley number & $\num{9.74d-2}$ &\\
		$\mathit{Sc}$ & Schmidt number & $\num{1282.63}$ &
	\end{tabular}
	\caption{
		Dimensional and dimensionless parameters in \cref{eq_dim_var,eq_dim_var_vel} for the transport of illuminated fluorescein dye in an oscillatory Couette flow of water.
		The parameters are mainly taken from the experimental study by \citet{ding_enhanced_2021}.
		\label{tab_parameters}
	}
\end{table}

\begin{figure}
	\centering
	{\includegraphics{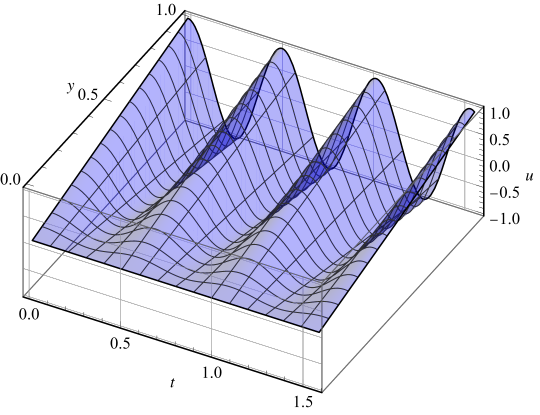}}
	\caption{
		Dimensionless velocity profile $u(y,t)$ of the oscillatory Couette flow with the parameters listed in \cref{tab_parameters}.
		\label{velocity_profile}}
\end{figure}

Following the experimental study by \citet{ding_enhanced_2021}, we consider the transport problem of illuminated fluorescein dye in water.
We perform the numerical simulation and analytical investigation based on the parameters listed in \cref{tab_parameters}.
The temporal evolution of the velocity profile is shown in \cref{velocity_profile}.
Due to the small Womersley number, the velocity profile is almost linear along the channel, namely, the boundary layer effect is weak.

\subsection{Comparison with numerical simulation}
\label{sec_numerical_simulation}

To verify our analytical solution of the concentration moments, the Brownian dynamics simulation method (also known as the Monte Carlo method) is adopted to obtain the numerical solutions.
The algorithm can be found in the recent paper by \citet[\S3]{ding_enhanced_2021} (see also \citet[Appendix A]{wang_vertical_2021}; \citet[Appendix B]{jiang_transient_2021}).

First, we need to find the corresponding stochastic differential equations (SDEs) for the original transport problem \labelcref{eq_problem_con}:
\begin{subequations}
\label{eq_SDEs}
\begin{align}
  \mathrm{d} X_t &= u (Y_t, t) \; \mathrm{d} t + \frac{\sqrt{2}}{\mathit{Pe}} \; \mathrm{d} W_1,
  \\
  \mathrm{d} Y_t &= \sqrt{2}\; \mathrm{d} W_2,
\end{align}
\end{subequations}
where $X_t$ and $Y_t$ are the random horizontal and transverse coordinates of the position of a particle, respectively. 
$W_1$ and $W_2$ are independent standard Brownian motions.

\begin{figure}
	\centering
	{\includegraphics{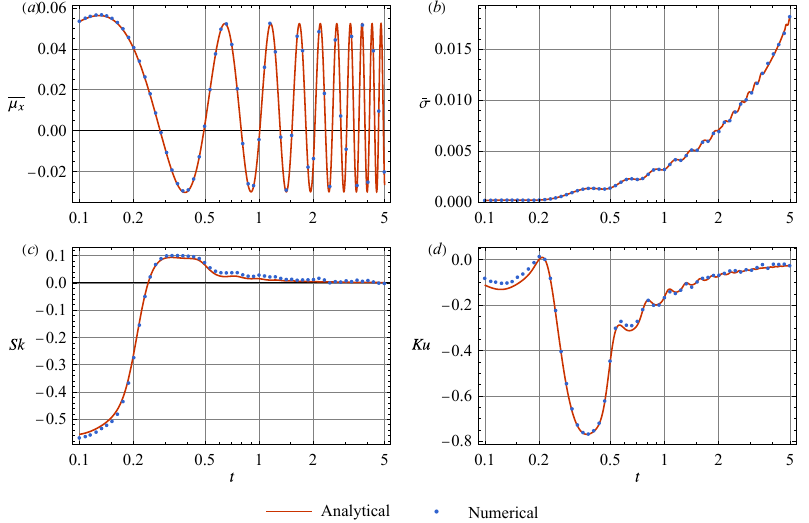}}
	\caption{
	Comparison between the analytical solution of concentration moments and the numerical result by Brownian dynamics simulation: (\textit{a}) the first order moment of mean concentration distribution; 
	(\textit{b}) the variance of $\bar{C}$;
	(\textit{c}) the skewness of $\bar{C}$;
	and (\textit{d}) the kurtosis of $\bar{C}$.
	The initial condition is a point-source release at $y_0 = 0.75$.
	Other parameters are listed in \cref{tab_parameters}.
	\label{numerical_comparison_bd}}
\end{figure}

In our simulation, the Euler--Maruyama method is adopted to numerically solve the SDEs.
The details are given in \cref{sec_num_sde}.
After checking the independence of the time step ($\Delta t$) and the number of particles ($N$), we take $\Delta t = 10^{-3}$ and $N = 10^5$.
For the point-source release considered in \cref{numerical_comparison_bd}, all particles are initially released at $x_0=0$ and $y_0 = 0.75$ (which corresponds to $C_{{\text{ini}}} = \delta(y-0.75)$ in \cref{eq_init_C}).
The trajectories of particles are recorded.
The positions of particles are used to statistically calculate concentration moments.

For the analytical solution, the infinite summations in the solution expressions such as \cref{eq_solution_P0} will be truncated for calculation.
After checking the independence of the number of function basis \labelcref{eq_eigenfunction}, we take the first eight eigenfunctions of $y$.
For the Fourier series of $t_1$ for the oscillation, we take $| k | \leqslant 4$ because we only consider concentration moments up to the fourth order for the discussion of kurtosis and the velocity profile is of one frequency.
Higher-order moments will require more Fourier components of $t_1$.

As shown in \cref{numerical_comparison_bd}, the analytical solutions of the transport characteristics of the mean concentration distribution $\overline{C}$ agree well with the numerical results, showing the correctness of the proposed auxiliary-time extension procedure.
Because we consider a point-source release as the initial condition, the truncation error may be large when $t$ is small.
More eigenfunctions may be required for the higher-order moments.
The linear--log plot (with a logarithmic scale of time) is used to highlight the alignment of curves in the initial short-time period, showing that the current configurations for the analytical and numerical solution procedures are sufficient.

\section{Effect of point-source release}
\label{sec_initial_condition}

With the auxiliary-time extension method being verified, one can investigate the effects of transport and flow parameters such as frequency, Schmidt number, Womersley number, etc.
These basic effects have been discussed in many previous studies \citep{chatwin_longitudinal_1975,vedel_transient_2012,ding_enhanced_2021,singh_significance_2023}.
In this study, our focus is the effects of the point-source release and the phase shift of velocity (see \cref{sec_phase_shift}) on the transport characteristics (especially the higher-order moments including skewness and kurtosis), which can also highlight the tractability of the auxiliary-time extension method.

In this section, we show the generality of the solution procedure by considering different initial conditions.
The generality derives from Barton's general solution expressions \citep{barton_method_1983} because they are based on an arbitrary initial condition and thus have wide applicability.
We consider the same oscillatory Couette flow and parameters as in \cref{sec_numerical_simulation} (see \cref{tab_parameters}).
Both the point-source release ($C_{{\text{ini}}} = \delta(y-y_0)$) and the uniform release ($C_{{\text{ini}}} = 1$) are considered.
For the point-source case, three different initial positions $y_0 = 0$, $y_0 = 0.5$, and $y_0 = 1$ are considered.

\begin{figure}
	\centering
	{\includegraphics{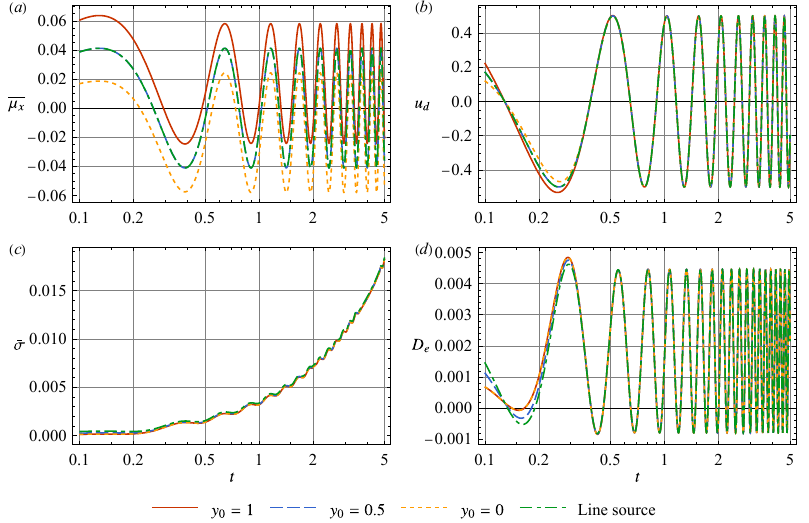}}
	\caption{Temporal evolution of (\textit{a}) the mean longitudinal position, (\textit{b}) the effective drift velocity, (\textit{c}) the  variance of mean concentration distribution, and (\textit{d}) the effective diffusion coefficient for different initial conditions.
	$y_0$ is the release position of the point-source release while "Line source" denotes the uniform release.
	The parameters are listed in \cref{tab_parameters}.
	\label{initial_condition_drift_disper}}
\end{figure}

First, we discuss the effect of point-source release position on the first-order moment of the mean concentration distribution, as shown in \cref{initial_condition_drift_disper}(\textit{a,b}).
Obviously, the mean position of the solute cloud is strongly affected by the initial condition, while the effective drift velocity is weakly affected.
Because the absolute value of the velocity at the upper wall is larger than that near the bottom wall, the effective drift velocity of the case with $y_0=1$ is larger than that of the case with $y_0=0$.
The curves of $u_d$ quickly overlap with each other after a short time (about half a period).
Note that $u_d$ of the case with $y_0=0.5$ and the case with a transversely uniform release are nearly the same. 
This is because the Womersley number is so small that the velocity profile is nearly linear along the transverse direction, as shown in \cref{velocity_profile}.
$u_d$ is the solute-weighted average velocity.
The transport problem with a line-source release is nearly symmetric with respect to the centerline of the channel.
Thus, its $u_d$ is nearly equal to the velocity value at the centerline, which is similar to the point-source case with $y_0=0.5$.

For the second-order concentration moment, the temporal evolution of the variance of $\bar{C}$ and the effective diffusion coefficient is shown in \cref{initial_condition_drift_disper}(\textit{c,d}).
There is no doubt that the source-release form will affect these characteristics in the initial stage \citep{vedel_time-dependent_2014}.
Similar to the effective drift velocity, $D_e$ of all the cases are nearly the same after a short time because the angular frequency $\omega$ is large.
Note that $D_e$ of the case with $y_0=0$ is nearly the same as that of the case with $y_0=1$.
This is because of the quasi-linear velocity profile (with a small $\mathit{Wo}$).
The spread of the solute cloud relative to the mean position is nearly symmetric for these two cases (or imagine the opposite case with an oscillating bottom wall instead of the upper wall).
Besides, it is well known that $D_e$ can be negative in the transport problem with oscillatory flows \citep{chatwin_longitudinal_1975,smith_contaminant_1982, yasuda_longitudinal_1984}, which can also be observed in \cref{initial_condition_drift_disper}(\textit{d}).
For the variance, all the curves are rising and are nearly overlapped with each other as $t$ increases.

\begin{figure}
	\centering
	{\includegraphics{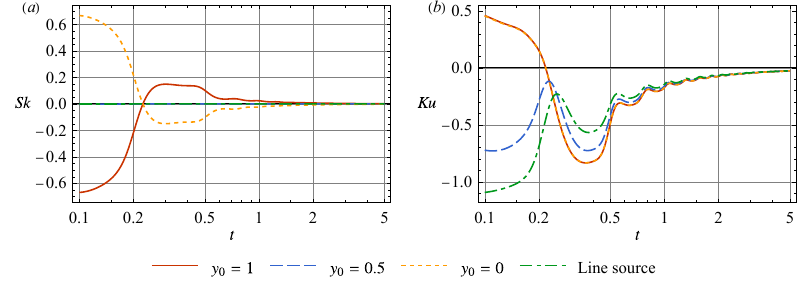}}
	\caption{
		Temporal evolution of (\textit{a}) skewness and (\textit{b}) kurtosis of the mean concentration distribution for different initial conditions.
		$y_0$ is the release position of the point-source release while "Line source" denotes the uniform release.
		Other parameters are listed in \cref{tab_parameters}.
		\label{initial_condition_skew_kurt}}
\end{figure}

The results for the higher-order moments of $\bar{C}$, skewness and kurtosis, are shown in \cref{initial_condition_skew_kurt}.
Similar to the case with a steady flow, both the skewness and kurtosis of the oscillatory case will still approach zero as $t$ increases because the Taylor dispersion regime is reached and $\bar{C}$ becomes Gaussian.
Evidently, the temporal evolution of skewness and kurtosis will be strongly affected by the initial condition.

For skewness, the point-source case with $y_0=0.5$ and the line-source case have a nearly zero value of $\mathit{Sk}$, which means that $\bar{C}$ is nearly symmetric with respect to the mean position of the solute cloud.
This is because the velocity profile is nearly linear and the transport problem is quasi-symmetric.
For the point-source cases with $y_0=0$ and $y_0=1$, 
their skewness values are nearly equal in magnitude but opposite in sign due to the quasi-symmetry of this specific transport problem.
A positive $\mathit{Sk}$ means that the distribution is upstream-skewed and vice versa.
Note that the sign of skewness can be changed during the transport process due to the oscillation of the velocity profile.

The transient evolution of kurtosis is much more complicated, although the oscillation induced by the flow is not apparent.
For the point-source cases with $y_0=0$ and $y_0=1$,
their kurtosis curves nearly overlap with each other due to the quasi-symmetry of this specific transport problem with a small $\mathit{Wo}$.
When the time is small, the tailedness is larger than the Gaussian distribution (positive $\mathit{Ku}$, leptokurtic).
However, the sign of kurtosis is changed during the transport process due to the velocity oscillation.
For the other two cases, the absolute value of kurtosis of the line-source case is larger than that of a point source with $y_0=0.5$.
But this trend can be reversed as time increases.

\section{Effect of phase shift of velocity profile}
\label{sec_phase_shift}

Another benefit of the auxiliary-time extension method is that it can easily deal with the phase shift effect of the velocity field on the transport characteristics.
\citet{ding_enhanced_2021} have discussed this effect on skewness's sign change, with the transversely uniform release.
Here, we further investigate the effect and show the result of other characteristics.

\subsection{Solution procedure to address phase shift}
\label{sec_phase_shift_solution}

First, we follow \citet{ding_enhanced_2021} and add a phase shift $s$ in the oscillatory velocity field as
\begin{equation}
  u_s (y, t) \triangleq u (y, t + s).
\end{equation}
Then replace the original velocity field $u$ in \cref{eq_concen} with the shifted velocity field $u_s$.

Typically, one needs to re-solve the whole problem of the concentration moments to impose the phase shift.
But with the auxiliary-time extension method, we can directly obtain the solution with a simple transformation of the time variable $t_1$.
Notice that $t_1$ is imposed in \cref{eq_t_0_and_t_1} as $t_1 = \omega t$.
To address the phase shift, we impose the following time variables instead:
\begin{equation}
	t_0 = t, \quad t_{1 s} = \omega (t + s).
\end{equation}
Notice that $t_{1 s} = t_1 + \omega s$.
Then, similar to \cref{eq_u_1}, we define $u_{s1}$ to replace the oscillatory velocity field for the shifted two-time-variable system and 
\begin{align}
\notag
	u_{s1} (y, t_{1 s}) & \triangleq u_s (y, t) = u (y, t + s)
\\ \notag
	&	= u (y, \frac{t_{1 s}}{\omega}-s + s) = u (y, \frac{t_{1 s}}{\omega})
\\ 
	&	= u_1 (y, t_{1 s}).
\end{align}
Namely, the function $u_{s1}$ is in the same form as $u_1$.
Thus, the governing equation for the shifted two-time-variable system is the same as that of the original system in \cref{eq_govern_P}.

To sum up, there is no need to re-solve the whole problem for the phase shift in flow velocity. 
The solution of the moments for the shifted two-time-variable system can be obtained by replacing $t_1$ with $t_{1 s}$ in the solution of the original two-time-variable system given in \cref{sec_solution_moments}.
Namely, 
\begin{equation}
	P_{s n} (y, t_0, t_{1 s}) = P_n (y, t_0, t_{1 s}),
\end{equation}
where $P_{s n}$ is the $n$-th order moment of the shifted two-time-variable system.
Based on \cref{eq_C_and_P_moment}, the $n$-th order concentration moments $C_{s n}$ of the shifted problem is
\begin{equation}
	\label{eq_C_and_P_moment_phase_shift}
	C_{s n}(y, t) = P_{s n} (y, t_0, t_{1 s}) = P_n (y, t, \omega (t + s)).
\end{equation}

Note that one could not obtain $C_{s n}(y,t)$ through $C_{n}(y,t)$ by simply replacing $t$ with $t + s$.
Fortunately, this could be done by $ P_n$ with $t_1$ replaced by $t_1 + \omega s$, which can be considered as a benefit of the auxiliary-time extension method.
Besides, because $t_1$ is only related to the oscillatory part, the decay rate to the Taylor dispersion regime is not affected by the phase shift.

We also perform the numerical verification for the phase shift effect, as shown in \cref{numerical_comparison_bd_phase_shift}.
The analytical result by \cref{eq_C_and_P_moment_phase_shift} agrees well with the numerical result.

\subsection{Results of transport characteristics}

To illustrate the effect of the phase shift on the transport characteristics, we consider the same oscillatory velocity profile as that in previous sections, but with different phase shift $s=0, \pi/4, \pi, \pi/2, 3\pi/4$.
A point-source release is considered ($y_0 = 1$).
Note that the case with $s=\pi$ is the reverse case of the case with $s=0$, due to the symmetry of the transport problem.
Thus, only four cases are considered.

\begin{figure}
	\centering
	{\includegraphics{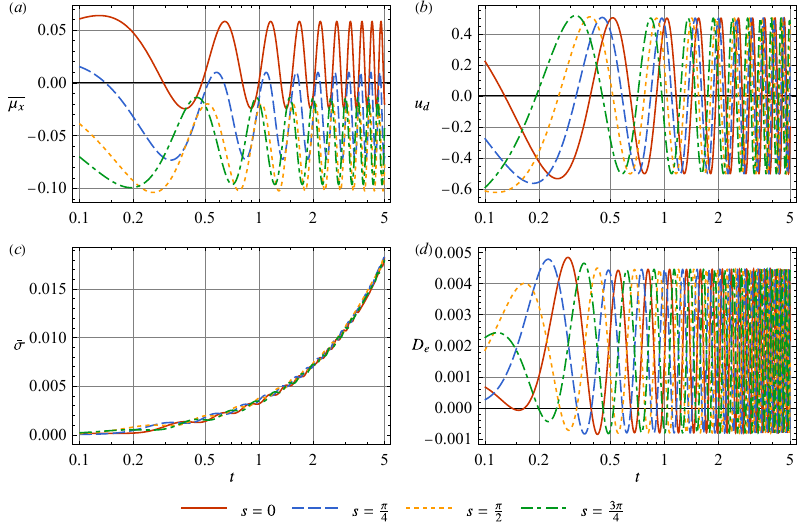}}
	\caption{
	Temporal evolution of (\textit{a}) the mean longitudinal position, (\textit{b}) the effective drift velocity, (\textit{c}) the  variance of mean concentration distribution, and (\textit{d}) the effective diffusion coefficient for different phase shift $s$.
	The initial condition is a point-source release at $y_0 = 1$.
	Other parameters are listed in \cref{tab_parameters}.
	\label{phase_shift_drift_disper}}
\end{figure}

For the first-order moment of $\bar{C}$ and the effective drift, it can be seen in \cref{phase_shift_drift_disper}(\textit{a,b}) that the effect of the phase shift is remarkable.
Note that the effective drift is the solute-weighted average velocity while the overall transverse distribution is not affected by the velocity field.
Thus, the phase shift of the velocity will directly be reflected in the effective drift.
It then leads to the accumulative change in the mean position of the solute cloud as time increases.

\Cref{phase_shift_drift_disper}(\textit{c,d}) shows the phase shift effect on the variance of $\bar{C}$ and the effective diffusion coefficient.
Note that $D_e$ is initially zero because of the instantaneous release.
The phase shift will change the time when $D_e$ reaches its maximum value.
Then $D_e$ will decay to the oscillatory state and reach the Taylor dispersion regime.
For the variance, similar to the effect of the initial condition discussed in \cref{sec_initial_condition}, the accumulative influence of the phase shift is weak and all the curves are nearly close to each other as $t$ increases.

\begin{figure}
	\centering
	{\includegraphics{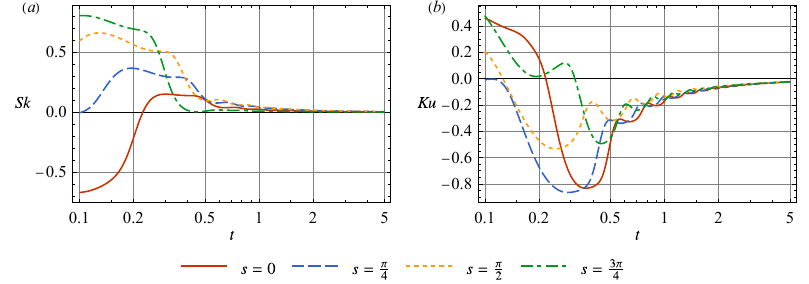}}
	\caption{
	Temporal evolution of (\textit{a}) skewness and (\textit{b}) kurtosis of the mean concentration distribution for different phase shift $s$.
	The initial condition is a point-source release at $y_0 = 1$.
	Other parameters are listed in \cref{tab_parameters}.
	\label{phase_shift_skew_kurt}}
\end{figure}

As discussed by \citet{ding_enhanced_2021}, the phase shift can change the sign of skewness, 
which is clearly observed in \cref{phase_shift_skew_kurt}(\textit{a}).
Unlike the line-source case in \citet{ding_enhanced_2021}, the point-source case discussed here has a large non-zero skewness in the initial stage of the transport process. 
Thus, the oscillation of skewness is not apparent.
However, the phase shift of the velocity field can still make a qualitative change in the asymmetry of the mean concentration distribution.
If we compare the cases with $s=0$ and $s=\pi$ (not shown here), then the curves will be flipped with respect to the $t$-axis due to the symmetric property of the transport problem.

For kurtosis, the phase shift will change the sign of $\mathit{Ku}$, as shown in \cref{phase_shift_skew_kurt}(\textit{b}).
The phase shift of the velocity field can be transferred to kurtosis, as the oscillation of $\mathit{Ku}$ is apparent during the initial stage of the transport process.
For the four considered cases, the temporal evolution curves have many intersections with each other even in the first period, showing the phase shift effect is complex.

\section{Conclusions} \label{sec_conclusions}

Understanding the transient dispersion in oscillatory flows is crucial for many practical applications.
The method of concentration moments is a powerful analytical tool for analyzing the transport characteristics and has been well developed for the transport problems in steady flows.
\citet{barton_method_1983} has provided the general solutions for the first-few-order moments, using  separation of variables.
However, for unsteady flows, these results cannot be applied directly.

Many previous studies \citep{mukherjee_dispersion_1988,vedel_transient_2012}
had to use the moment method from the very beginning: re-solving the governing equation of moments.
However, due to the complication induced by the time-periodic velocity field, the solution procedure is not as straightforward and general as that by \citet{barton_method_1983}.
For higher-order statistics, the skewness has been derived only for a specific initial condition \citep{ding_enhanced_2021}, while the solution for kurtosis remains analytically intractable, necessitating the use of numerical methods \citep{singh_significance_2023}.

This manuscript proposes a novel analytical procedure by the introduction of an auxiliary oscillation time variable for the inherent oscillatory characteristics of the dispersion and extends the original transport problem to a two-time-variable problem, with
the new time dimension virtually viewed as an additional ``transverse'' space.
Then there is no need to re-solve the moment equation: the general solution expressions by \citet{barton_method_1983} now can be used directly because the flow velocity becomes ``steady'' after splitting the time variable.
We remark that although the introduction of the time variables bears structural similarities to the two-time-scale homogenization method,
these two methods are fundamentally distinct in their underlying principles.
The present method does not rely on either a small perturbation parameter or scale-separation analysis.
The objective here is the concentration moments, not the long-time asymptotic concentration distribution (as in the homogenization method).

The solution procedure for concentration moments is then greatly simplified to only two steps: (i) identifying the eigenvalue problem and (ii) substituting the eigenvalues and eigenfunctions in Barton's expressions.
This auxiliary-time extension treatment not only offers an intuitive physical perspective for the velocity oscillation but also clarifies the solution structure of concentration moments.
The complications encountered in previous studies can be avoided, making it feasible to investigate the temporal evolution of higher-order statistics.

As a preliminary verification, we examine the transport problem in an oscillatory Couette flow.
The numerical results by the Brownian dynamics simulation method are used for comparison and show good agreement with the analytical solutions by the auxiliary-time extension method.

Furthermore, to highlight the tractability of the auxiliary-time extension method, we investigate the effects of the point-source release and the phase shift of velocity on the transport characteristics, especially on higher-order statistics like skewness and kurtosis.
The generality of the solution procedure derives from Barton's general solution expressions for arbitrary initial conditions.
For the phase shift of velocity, a simple transformation of the newly introduced time variable can be used to handle the phase shift effect without re-solving the moment equation.
The oscillation of skewness is not apparent in the case with a point-source release.
The phase shift of the velocity field can be remarkably transferred to the temporal evolution of kurtosis.
Both these two effects can change the sign of skewness and kurtosis, qualitatively affecting the asymmetry and the tailedness of the mean concentration distribution.


\backsection[Acknowledgements]{
The authors would like to thank Dr Bohan Wang for constructive discussions.
}

\backsection[Funding]{
	This work is supported by the Science and Technology Development Fund (FDCT) of Macao SAR (Project Nos.\ 0018/2023/ITP1, 0025/2024/RIA1, 0031/2024/AMJ, and 0001/2024/NRP) and the National Natural Science Foundation of China (Grant No.\ 52079001).
}

\backsection[Declaration of interests]{The authors report no conflict of interest.}


\backsection[Author ORCID]{
	\urlstyle{same}
	\\
	Weiquan Jiang \url{https://orcid.org/0000-0002-2528-7736};
	\\
	Guoqian Chen \url{https://orcid.org/0000-0003-1173-6796}.
}


\appendix

\section{Governing equations of moments}
\label{sec_eqs_moments}

To derive the governing equation of concentration moments, it is assumed that concentration decays exponentially (cf.\ \citet{aris_dispersion_1956}).
Integrating \cref{eq_concen} according to the definition of $C_n$ in \cref{eq_def_con_moment}, 
and using the integration by parts formula, we have
\begin{equation} \label{eq_govern_Cn}
	\frac{\partial C_n}{\partial t}
    - \frac{\partial^2 C_n}{\partial y^2} 
     = n \, u(y, t) C_{n - 1} + \frac{n (n-1)}{\mathit{Pe}^2}C_{n - 2},
\end{equation}
for $n=0,1,2,\ldots$.
Notice that $C_{-1}=C_{-2}=0$ are auxiliary terms.
The boundary conditions of $C_n$ are the same as that of $C$ in \cref{eq_bc_C}:
\begin{equation}
	\left. \frac{\partial C_n}{\partial y} \right|_{y = 0} 
      = \left. \frac{\partial C_n}{\partial y} \right|_{y = 1} = 0.
\end{equation}
For the initial condition, according to \cref{eq_init_C}, we have
\begin{equation}
	C_n | _{t = 0}  = 
      \left\{\begin{array}{ll}
        C_{{\text{ini}}} (y), & n = 0\\
        0 , & n = 1, 2, \ldots.
      \end{array} \right.
\end{equation}

Note that the governing equation \labelcref{eq_govern_Cn} of $C_n$ contains the time-dependent velocity field $u(y,t)$ as a source term on the left side.
Thus, obtaining time-integral solutions for $C_n$ is more complicated compared to steady flow cases, particularly for higher-order moments.

\section{Solution of second-order moments}
\label{sec_solution_moments_2_and_higher}

For the second-order moment, based on equation (2.14) by \citet{barton_asymptotic_1984}, the solution is
\begin{align} \label{eq_solution_M2}
	\notag
	{{P}}_2 (y, t_0, t_1)
	= & \sum^{\infty}_{i=0} a_i B_{i, i}^2 t_0^2  \mathrm{e}^{- \lambda_i t_0} {f}_i
	+ \sum^{\infty}_{\substack{i,j=0 \\ j\neq i}} \frac{2 E_{i, j} a_j (\mathrm{e}^{- \lambda_j t_0} - \mathrm{e}^{- \lambda_i t_0}) }{\lambda_i - \lambda_j} {f}_i
	+\sum^{\infty}_{i=0} 2 a_i E_{i,i}  t_0 \mathrm{e}^{-\lambda_i t_0}  f_i 
	\\ \notag
	&+\sum^{\infty}_{\substack{i,j,k=0 \\ k\neq j, k\neq i}} 
		\frac{2 a_k B_{j, k} B_{i, j} (\mathrm{e}^{- \lambda_k t_0} - \mathrm{e}^{-\lambda_i t_0}) }{(\lambda_k - \lambda_i) (\lambda_k - \lambda_j) }{f}_i
	+\sum^{\infty}_{\substack{i,j=0 \\ i\neq j}} 
		\frac{2 a_i B_{j, i} B_{i, j} t_0 \mathrm{e}^{- \lambda_i t_0} }{\lambda_j -\lambda_i} {f}_i
	\\ \notag
	&+\sum^{\infty}_{\substack{i,j,k=0 \\ k\neq j,  j\neq i}} 
		\frac{2 a_k B_{j, k} B_{i, j} (\mathrm{e}^{- \lambda_j t_0} - \mathrm{e}^{-\lambda_i t_0}) }{(\lambda_k - \lambda_j)  (\lambda_i - \lambda_j)}{f}_i
	+\sum^{\infty}_{\substack{i,k=0 \\ k\neq i}} 
		\frac{2 a_k B_{i, k} B_{i, i} t_0 \mathrm{e}^{- \lambda_i t_0} }{\lambda_k -\lambda_i} {f}_i
	\\
	&+\sum^{\infty}_{\substack{i,j=0 \\ j\neq i}} 
		\frac{2 a_j B_{j, j} B_{i, j}  [\mathrm{e}^{- \lambda_i t_0} - \mathrm{e}^{-\lambda_j t_0} + (\lambda_i - \lambda_j) t_0 \mathrm{e}^{- \lambda_j t_0}]}{(\lambda_j - \lambda_i)^2} {f}_i
		,
\end{align}
where $E_{i, j}$ is the coefficient related to the diffusion coefficient $D$,
\begin{align}
	E_{i, j} &\triangleq \langle {f}_i,  D {f}_j \rangle
	\\ \notag
	&=  \int^{2 \pi}_0 \int^1_0 f_i^{\star} (y, t_1) D f_j (y, t_1) \; \mathrm{d} y \mathrm{d} t_1 
	\\ \notag
	&= D \delta_{i j},
\end{align}
for $i, j = 0, 1, \ldots$.
Note that $D=1/\mathit{Pe}^2$ for the transport problem considered in \cref{eq_govern_P}.

For the third and fourth-order moments, the readers can refer to equation (3.17) of \citet{barton_method_1983}, an analytical expression of the third-order moment can also be found in Appendix A of \citet{yang_migration_2021}.
Their expressions are too long to be reproduced here.
For practical purposes, readers can refer to the study of \citet{jiang_analytical_2022}, with a Mathematica notebook provided at \url{https://zenodo.org/doi/10.5281/zenodo.14927459}.

\section{Numerical stochastic differential equations}
\label{sec_num_sde}

\subsection{Euler--Maruyama scheme}

We use the Euler--Maruyama scheme to solve the SDEs \labelcref{eq_SDEs}.
Let $X_n$ and $Y_n$ denote the position of a particle at the $n$-th time step $t_n=n\Delta t$.
We have 
\begin{subequations}
	\label{eq_Euler_Maruyama}
	\begin{align}
	X_{n + 1} &= X_n + u (Y_n, t_n) \Delta t + \frac{\sqrt{2 \Delta t}}{\mathit{Pe}} \Delta W_1,
\\
	Y_{n + 1} & = Y_n + \sqrt{2 \Delta t} \Delta W_2,
	\end{align}
\end{subequations}
where $\Delta W_1$ and $\Delta W_2$ are random variables generated at each time step.
They satisfy the standard normal distribution, i.e., $\Delta W_1 \sim \mathrm{N}(0, 1)$ and $\Delta W_2 \sim \mathrm{N}(0, 1)$.
$\mathrm{N}$ denotes the normal distribution.
In our code, we add $t_{n+1}=t_n + \Delta t$ to \cref{eq_Euler_Maruyama} such that the whole system is autonomous.

On the channel walls ($y=0$ and $y=1$), the specular reflection model (billiard-like reflection) is adopted to simulate the non-penetration boundary condition \labelcref{eq_bc_C}.
If a particle exceeds the boundaries after the calculation at the $n+1$ time step, it is reflected back to the channel:
\begin{subequations}
\begin{align}
	Y_{n + 1} \rightarrow - Y_{n + 1}, \quad \text{if} \quad Y_{n + 1} < 0,
\\
	Y_{n + 1} \rightarrow 2 - Y_{n + 1}, \quad \text{if} \quad Y_{n + 1} > 1.
\end{align}
\end{subequations}

For the initial condition, we set $X_0 = 0$ and $Y_0 = 0.75$ for a point-source release at $y_0 = 0.75$.
For the uniform release ($C_{{\text{ini}}} = 1$), $X_0$ is a random variable generated from the uniform distribution on the interval $[0, 1]$.

\subsection{Phase shift}

\begin{figure}
	\centering
	{\includegraphics{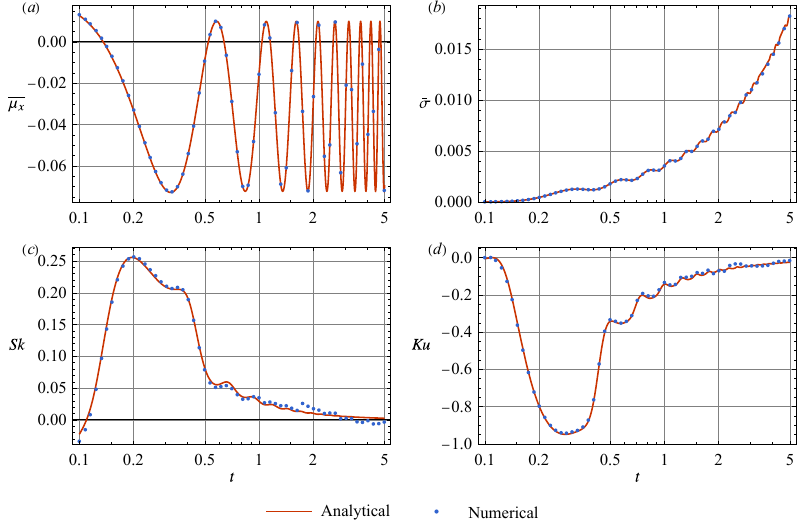}}
	\caption{
	Comparison between the analytical solution of concentration moments and the numerical result for the phase shift effect: (\textit{a}) the first order moment of mean concentration distribution; 
	(\textit{b}) the variance of $\bar{C}$;
	(\textit{c}) the skewness of $\bar{C}$;
	and (\textit{d}) the kurtosis of $\bar{C}$.
	The initial condition is a point-source release at $y_0 = 1$.
	Other parameters are listed in \cref{tab_parameters}.
	\label{numerical_comparison_bd_phase_shift}
	}
\end{figure}

We also perform the numerical simulation for the case with a phase shift of the velocity field.
\Cref{numerical_comparison_bd_phase_shift} shows the numerical verification for the treatment of the phase shift effect presented in \cref{sec_phase_shift_solution}.
The analytical solutions of the transport characteristics agree well with the numerical results.

\FloatBarrier
\bibliographystyle{jfm} 
\bibliography{lib_oscillatory}

\begin{thebibliography}{79}
\expandafter\ifx\csname natexlab\endcsname\relax\def\natexlab#1{#1}\fi
\def\au#1{#1} \def\ed#1{#1} \def\yr#1{#1}\def\at#1{#1}\def\jt#1{\textit{#1}}
  \def\bt#1{#1}\def\bvol#1{\textbf{#1}} \def\vol#1{#1} \def\pg#1{#1}
  \def\publ#1{#1}\def\arxiv#1{#1}\def\org#1{#1}\def\st#1{\textit{#1}}

\bibitem[Allen(1982)]{allen_numerical_1982}
{\sc \au{Allen, C.~M.}} \yr{1982}  \at{Numerical simulation of contaminant
  dispersion in estuary flows}.  \jt{Proc. R. Soc. Lond. A}  \bvol{381}~(1780),
   \pg{179--194}.

\bibitem[Aminian {\em et~al.\/}(2016)Aminian, Bernardi, Camassa, Harris \&
  McLaughlin]{aminian_how_2016}
{\sc \au{Aminian, M.}, \au{Bernardi, F.}, \au{Camassa, R.}, \au{Harris, D.~M.}
  \& \au{McLaughlin, R.~M.}} \yr{2016}  \at{How boundaries shape chemical
  delivery in microfluidics}.  \jt{Science}  \bvol{354}~(6317),
  \pg{1252--1256}.

\bibitem[Aminian {\em et~al.\/}(2015)Aminian, Bernardi, Camassa \&
  McLaughlin]{aminian_squaring_2015}
{\sc \au{Aminian, M.}, \au{Bernardi, F.}, \au{Camassa, R.} \& \au{McLaughlin,
  R.~M.}} \yr{2015}  \at{Squaring the circle: geometric skewness and symmetry
  breaking for passive scalar transport in ducts and pipes}.  \jt{Phys. Rev.
  Lett.}  \bvol{115}~(15),  \pg{154503}.

\bibitem[Andersson \& Berglin(1981)]{andersson_dispersion_1981}
{\sc \au{Andersson, B.} \& \au{Berglin, T.}} \yr{1981}  \at{Dispersion in
  laminar flow through a circular tube}.  \jt{Proc. R. Soc. Lond. A}
  \bvol{377}~(1770),  \pg{251--268}.

\bibitem[Aquino(2024)]{aquino_equilibrium_2024}
{\sc \au{Aquino, T.}} \yr{2024}  \at{Equilibrium distributions under
  advection--diffusion in laminar channel flow with partially absorbing
  boundaries}.  \jt{J. Fluid Mech.}  \bvol{985},  \pg{A16}.

\bibitem[Aris(1956)]{aris_dispersion_1956}
{\sc \au{Aris, R.}} \yr{1956}  \at{On the dispersion of a solute in a fluid
  flowing through a tube}.  \jt{Proc. R. Soc. Lond. A}  \bvol{235}~(1200),
  \pg{67--77}.

\bibitem[Aris(1960)]{aris_dispersion_1960}
{\sc \au{Aris, R.}} \yr{1960}  \at{On the dispersion of a solute in pulsating
  flow through a tube}.  \jt{Proc. R. Soc. Lond. A}  \bvol{259}~(1298),
  \pg{370--376}.

\bibitem[Bandyopadhyay \& Mazumder(1999)]{bandyopadhyay_unsteady_1999}
{\sc \au{Bandyopadhyay, S.} \& \au{Mazumder, B.~S.}} \yr{1999}  \at{Unsteady
  convective diffusion in a pulsatile flow through a channel}.  \jt{Acta Mech.}
   \bvol{134}~(1-2),  \pg{1--16}.

\bibitem[Barik \& Dalal(2019)]{barik_multi-scale_2019}
{\sc \au{Barik, S.} \& \au{Dalal, D.~C.}} \yr{2019}  \at{Multi-scale analysis
  for concentration distribution in an oscillatory {{Couette}} flow}.
  \jt{Proc. R. Soc. A}  \bvol{475}~(2221),  \pg{20180483}.

\bibitem[Barton(1983)]{barton_method_1983}
{\sc \au{Barton, N.~G.}} \yr{1983}  \at{On the method of moments for solute
  dispersion}.  \jt{J. Fluid Mech.}  \bvol{126},  \pg{205--218}.

\bibitem[Barton(1984)]{barton_asymptotic_1984}
{\sc \au{Barton, N.~G.}} \yr{1984}  \at{An asymptotic theory for dispersion of
  reactive contaminants in parallel flow}.  \jt{J. Aust. Math. Soc. Ser. B
  Appl. Math.}  \bvol{25}~(3),  \pg{287--310}.

\bibitem[Bowden(1965)]{bowden_horizontal_1965}
{\sc \au{Bowden, K.~F.}} \yr{1965}  \at{Horizontal mixing in the sea due to a
  shearing current}.  \jt{J. Fluid Mech.}  \bvol{21}~(1),  \pg{83--95}.

\bibitem[Brenner \& Edwards(1993)]{brenner_macrotransport_1993}
{\sc \au{Brenner, H.} \& \au{Edwards, D.~A.}} \yr{1993} {\em Macrotransport
  {{Processes}}\/}.  \publ{Stoneham: Butterworth-Heinemann}.

\bibitem[Camassa {\em et~al.\/}(2010)Camassa, Lin \&
  McLaughlin]{camassa_exact_2010}
{\sc \au{Camassa, R.}, \au{Lin, Z.} \& \au{McLaughlin, R.~M.}} \yr{2010}
  \at{The exact evolution of the scalar variance in pipe and channel flow}.
  \jt{Commun. Math. Sci.}  \bvol{8}~(2),  \pg{601--626}.

\bibitem[Chatwin(1970)]{chatwin_approach_1970}
{\sc \au{Chatwin, P.~C.}} \yr{1970}  \at{The approach to normality of the
  concentration distribution of a solute in a solvent flowing along a straight
  pipe}.  \jt{J. Fluid Mech.}  \bvol{43}~(2),  \pg{321--352}.

\bibitem[Chatwin(1972)]{chatwin_cumulants_1972}
{\sc \au{Chatwin, P.~C.}} \yr{1972}  \at{The cumulants of the distribution of
  concentration of a solute dispersing in solvent flowing through a tube}.
  \jt{J. Fluid Mech.}  \bvol{51}~(1),  \pg{63--67}.

\bibitem[Chatwin(1975)]{chatwin_longitudinal_1975}
{\sc \au{Chatwin, P.~C.}} \yr{1975}  \at{On the longitudinal dispersion of
  passive contaminant in oscillatory flows in tubes}.  \jt{J. Fluid Mech.}
  \bvol{71}~(3),  \pg{513--527}.

\bibitem[Chu {\em et~al.\/}(2019)Chu, Garoff, Przybycien, Tilton \&
  Khair]{chu_dispersion_2019}
{\sc \au{Chu, H. C.~W.}, \au{Garoff, S.}, \au{Przybycien, T.~M.}, \au{Tilton,
  R.~D.} \& \au{Khair, A.~S.}} \yr{2019}  \at{Dispersion in steady and
  time-oscillatory two-dimensional flows through a parallel-plate channel}.
  \jt{Phys. Fluids}  \bvol{31}~(2),  \pg{022007}.

\bibitem[Das {\em et~al.\/}(2024)Das, Dhar, Kairi, Mondal \&
  Poddar]{das_analysis_2024}
{\sc \au{Das, D.}, \au{Dhar, S.}, \au{Kairi, R.~R.}, \au{Mondal, K.~K.} \&
  \au{Poddar, N.}} \yr{2024}  \at{Analysis of environmental transport of
  suspended sediment particles in a tidal wetland flow under the effect of
  floating vegetation absorption}.  \jt{Commun. Nonlinear Sci. Numer. Simul.}
  \bvol{132},  \pg{107888}.

\bibitem[Debnath {\em et~al.\/}(2019)Debnath, Saha, Mazumder \&
  Roy]{debnath_transport_2019}
{\sc \au{Debnath, S.}, \au{Saha, A.~K.}, \au{Mazumder, B.~S.} \& \au{Roy,
  A.~K.}} \yr{2019}  \at{Transport of a reactive solute in a pulsatile
  non-{{Newtonian}} liquid flowing through an annular pipe}.  \jt{J. Eng.
  Math.}  \bvol{116}~(1),  \pg{1--22}.

\bibitem[Dhar {\em et~al.\/}(2022)Dhar, Das, Poddar \&
  Mondal]{dhar_dispersion_2022}
{\sc \au{Dhar, S.}, \au{Das, D.}, \au{Poddar, N.} \& \au{Mondal, K.~K.}}
  \yr{2022}  \at{Dispersion of fine settling particles in a tidal wetland
  flow}.  \jt{J. Hydrol.}  \bvol{615},  \pg{128701}.

\bibitem[Ding(2023)]{ding_shear_2023}
{\sc \au{Ding, L.}} \yr{2023}  \at{Shear dispersion of multispecies electrolyte
  solutions in the channel domain}.  \jt{J. Fluid Mech.}  \bvol{970},
  \pg{A27}.

\bibitem[Ding {\em et~al.\/}(2021)Ding, Hunt, McLaughlin \&
  Woodie]{ding_enhanced_2021}
{\sc \au{Ding, L.}, \au{Hunt, R.}, \au{McLaughlin, R.~M.} \& \au{Woodie, H.}}
  \yr{2021}  \at{Enhanced diffusivity and skewness of a diffusing tracer in the
  presence of an oscillating wall}.  \jt{Res. Math. Sci.}  \bvol{8}~(3),
  \pg{34}.

\bibitem[Drazin \& Riley(2006)]{drazin_navier-stokes_2006}
{\sc \au{Drazin, P.~G.} \& \au{Riley, N.}} \yr{2006} {\em The {{Navier-Stokes
  Equations}}: {{A Classification}} of {{Flows}} and {{Exact Solutions}}\/}.
  \publ{Cambridge: Cambridge University Press}.

\bibitem[Fife \& Nicholes(1975)]{fife_dispersion_1975}
{\sc \au{Fife, P.~C.} \& \au{Nicholes, K. R.~K.}} \yr{1975}  \at{Dispersion in
  flow through small tubes}.  \jt{Proc. R. Soc. Lond. A}  \bvol{344}~(1636),
  \pg{131--145}.

\bibitem[Fischer(1972)]{fischer_mass_1972}
{\sc \au{Fischer, H.}} \yr{1972}  \at{Mass transport mechanisms in partially
  stratified estuaries}.  \jt{J. Fluid Mech.}  \bvol{53}~(4),  \pg{671--687}.

\bibitem[Frankel \& Brenner(1991)]{frankel_generalized_1991}
{\sc \au{Frankel, I.} \& \au{Brenner, H.}} \yr{1991}  \at{Generalized
  {{Taylor}} dispersion phenomena in unbounded homogeneous shear flows}.
  \jt{J. Fluid Mech.}  \bvol{230},  \pg{147--181}.

\bibitem[Gaver \& Grotberg(1990)]{gaver_dynamics_1990}
{\sc \au{Gaver, D.~P.} \& \au{Grotberg, J.~B.}} \yr{1990}  \at{The dynamics of
  a localized surfactant on a thin film}.  \jt{J. Fluid Mech.}  \bvol{213},
  \pg{127--148}.

\bibitem[Gill \& Sankarasubramanian(1971)]{gill_dispersion_1971}
{\sc \au{Gill, W.~N.} \& \au{Sankarasubramanian, R.}} \yr{1971}  \at{Dispersion
  of a non-uniform slug in time-dependent flow}.  \jt{Proc. R. Soc. Lond. A}
  \bvol{322}~(1548),  \pg{101--117}.

\bibitem[Grotberg \& Jensen(2004)]{grotberg_biofluid_2004}
{\sc \au{Grotberg, J.~B.} \& \au{Jensen, O.~E.}} \yr{2004}  \at{Biofluid
  mechanics in flexible tubes}.  \jt{Annu. Rev. Fluid Mech.}  \bvol{36},
  \pg{121--147}.

\bibitem[Grotberg {\em et~al.\/}(1990)Grotberg, Sheth \&
  Mockros]{grotberg_analysis_1990}
{\sc \au{Grotberg, J.~B.}, \au{Sheth, B.~V.} \& \au{Mockros, L.~F.}} \yr{1990}
  \at{An analysis of pollutant gas transport and absorption in pulmonary
  airways}.  \jt{J. Biomech. Eng.}  \bvol{112}~(2),  \pg{168--176}.

\bibitem[Guan \& Chen(2024)]{guan_streamwise_2024}
{\sc \au{Guan, M.} \& \au{Chen, G.}} \yr{2024}  \at{Streamwise dispersion of
  soluble matter in solvent flowing through a tube}.  \jt{J. Fluid Mech.}
  \bvol{980},  \pg{A33}.

\bibitem[Guan {\em et~al.\/}(2021)Guan, Zeng, Li, Guo, Wu \&
  Wang]{guan_transport_2021}
{\sc \au{Guan, M.~Y.}, \au{Zeng, L.}, \au{Li, C.~F.}, \au{Guo, X.~L.}, \au{Wu,
  Y.~H.} \& \au{Wang, P.}} \yr{2021}  \at{Transport model of active particles
  in a tidal wetland flow}.  \jt{J. Hydrol.}  \bvol{593},  \pg{125812}.

\bibitem[Hazra {\em et~al.\/}(1996)Hazra, Gupta \&
  Niyogi]{hazra_dispersion_1996}
{\sc \au{Hazra, S.~B.}, \au{Gupta, A.~S.} \& \au{Niyogi, P.}} \yr{1996}  \at{On
  the dispersion of a solute in oscillating flow through a channel}.  \jt{Heat
  Mass Transf.}  \bvol{31}~(4),  \pg{249--256}.

\bibitem[Holley {\em et~al.\/}(1970)Holley, Harleman \&
  Fischer]{holley_dispersion_1970}
{\sc \au{Holley, E.~R.}, \au{Harleman, D. R.~F.} \& \au{Fischer, H.~B.}}
  \yr{1970}  \at{Dispersion in homogeneous estuary flow}.  \jt{J. Hydraul.
  Div.}  \bvol{96}~(8),  \pg{1691--1709}.

\bibitem[Jensen {\em et~al.\/}(2009)Jensen, Rio, Hansen, Clanet \&
  Bohr]{jensen_osmotically_2009}
{\sc \au{Jensen, K.~H.}, \au{Rio, E.}, \au{Hansen, R.}, \au{Clanet, C.} \&
  \au{Bohr, T.}} \yr{2009}  \at{Osmotically driven pipe flows and their
  relation to sugar transport in plants}.  \jt{J. Fluid Mech.}  \bvol{636},
  \pg{371--396}.

\bibitem[Jiang \& Chen(2021)]{jiang_transient_2021}
{\sc \au{Jiang, W.} \& \au{Chen, G.}} \yr{2021}  \at{Transient dispersion
  process of active particles}.  \jt{J. Fluid Mech.}  \bvol{927},  \pg{A11}.

\bibitem[Jiang {\em et~al.\/}(2022)Jiang, Zeng, Fu \&
  Wu]{jiang_analytical_2022}
{\sc \au{Jiang, W.}, \au{Zeng, L.}, \au{Fu, X.} \& \au{Wu, Z.}} \yr{2022}
  \at{Analytical solutions for reactive shear dispersion with boundary
  adsorption and desorption}.  \jt{J. Fluid Mech.}  \bvol{947},  \pg{A37}.

\bibitem[Jiang \& Chen(2018)]{jiang_solution_2018}
{\sc \au{Jiang, W.~Q.} \& \au{Chen, G.~Q.}} \yr{2018}  \at{Solution of
  {{Gill}}'s generalized dispersion model: solute transport in {{Poiseuille}}
  flow with wall absorption}.  \jt{Int. J. Heat Mass Transf.}  \bvol{127},
  \pg{34--43}.

\bibitem[Jiang \& Grotberg(1993)]{jiang_bolus_1993}
{\sc \au{Jiang, Y.} \& \au{Grotberg, J.~B.}} \yr{1993}  \at{Bolus contaminant
  dispersion in oscillatory tube flow with conductive walls}.  \jt{J. Biomech.
  Eng.}  \bvol{115}~(4A),  \pg{424--431}.

\bibitem[Joshi {\em et~al.\/}(1983)Joshi, Kamm, Drazen \&
  Slutsky]{joshi_experimental_1983}
{\sc \au{Joshi, C.~H.}, \au{Kamm, R.~D.}, \au{Drazen, J.~M.} \& \au{Slutsky,
  A.~S.}} \yr{1983}  \at{An experimental study of gas exchange in laminar
  oscillatory flow}.  \jt{J. Fluid Mech.}  \bvol{133},  \pg{245--254}.

\bibitem[Karmakar {\em et~al.\/}(2023)Karmakar, Barik \&
  Raja~Sekhar]{karmakar_multi-scale_2023}
{\sc \au{Karmakar, T.}, \au{Barik, S.} \& \au{Raja~Sekhar, G.~P.}} \yr{2023}
  \at{Multi-scale analysis of concentration distribution in unsteady
  {{Couette}}--{{Poiseuille}} flows through a porous channel}.  \jt{Proc. R.
  Soc. A}  \bvol{479}~(2269),  \pg{20220494}.

\bibitem[Mazumder \& Das(1992)]{mazumder_effect_1992}
{\sc \au{Mazumder, B.~S.} \& \au{Das, S.~K.}} \yr{1992}  \at{Effect of boundary
  reaction on solute dispersion in pulsatile flow through a tube}.  \jt{J.
  Fluid Mech.}  \bvol{239},  \pg{523--549}.

\bibitem[Mederos {\em et~al.\/}(2020)Mederos, Arcos, Bautista \&
  M{\'e}ndez]{mederos_hydrodynamics_2020}
{\sc \au{Mederos, G.}, \au{Arcos, J.}, \au{Bautista, O.} \& \au{M{\'e}ndez,
  F.}} \yr{2020}  \at{Hydrodynamics rheological impact of an oscillatory
  electroosmotic flow on a mass transfer process in a microcapillary with a
  reversible wall reaction}.  \jt{Phys. Fluids}  \bvol{32}~(12),  \pg{122003}.

\bibitem[Mei(1992)]{mei_method_1992}
{\sc \au{Mei, C.~C.}} \yr{1992}  \at{Method of homogenization applied to
  dispersion in porous media}.  \jt{Transp. Porous Media}  \bvol{9}~(3),
  \pg{261--274}.

\bibitem[Morris(2001)]{morris_anomalous_2001}
{\sc \au{Morris, J.~F.}} \yr{2001}  \at{Anomalous migration in simulated
  oscillatory pressure-driven flow of a concentrated suspension}.  \jt{Phys.
  Fluids}  \bvol{13}~(9),  \pg{2457--2462}.

\bibitem[Mukherjee \& Mazumder(1988)]{mukherjee_dispersion_1988}
{\sc \au{Mukherjee, A.} \& \au{Mazumder, B.~S.}} \yr{1988}  \at{Dispersion of
  contaminant in oscillatory flows}.  \jt{Acta Mech.}  \bvol{74}~(1-4),
  \pg{107--122}.

\bibitem[Ng(2004)]{ng_time-varying_2004}
{\sc \au{Ng, C.-O.}} \yr{2004}  \at{A time-varying diffusivity model for shear
  dispersion in oscillatory channel flow}.  \jt{Fluid Dyn. Res.}
  \bvol{34}~(6),  \pg{335--355}.

\bibitem[Ng(2006)]{ng_dispersion_2006}
{\sc \au{Ng, C.-O.}} \yr{2006}  \at{Dispersion in steady and oscillatory flows
  through a tube with reversible and irreversible wall reactions}.  \jt{Proc.
  R. Soc. A}  \bvol{462}~(2066),  \pg{481--515}.

\bibitem[Pal {\em et~al.\/}(2025)Pal, Singh \& Murthy]{pal_solute_2025}
{\sc \au{Pal, P.~B.}, \au{Singh, S.} \& \au{Murthy, P. V. S.~N.}} \yr{2025}
  \at{Solute dispersion in unsteady and viscous flow regimes of a
  non-{{Newtonian}} fluid flow with periodic body acceleration/deceleration}.
  \jt{Phys. Fluids}  \bvol{37}~(2),  \pg{023113}.

\bibitem[Paul \& Mazumder(2008)]{paul_dispersion_2008}
{\sc \au{Paul, S.} \& \au{Mazumder, B.~S.}} \yr{2008}  \at{Dispersion in
  unsteady {{Couette}}--{{Poiseuille}} flows}.  \jt{Int. J. Eng. Sci.}
  \bvol{46}~(12),  \pg{1203--1217}.

\bibitem[Pavliotis(2002)]{pavliotis_homogenization_2002}
{\sc \au{Pavliotis, G.~A.}} \yr{2002}  \at{Homogenization theory for
  advection--diffusion equations with mean flow}. PhD thesis, Rensselaer
  Polytechnic Institute, Troy.

\bibitem[Pavliotis \& Stuart(2008)]{pavliotis_multiscale_2008}
{\sc \au{Pavliotis, G.~A.} \& \au{Stuart, A.}} \yr{2008} {\em Multiscale
  {{Methods}}: {{Averaging}} and {{Homogenization}}\/}.  \publ{New York:
  Springer}.

\bibitem[Pedley \& Kamm(1988)]{pedley_effect_1988}
{\sc \au{Pedley, T.~J.} \& \au{Kamm, R.~D.}} \yr{1988}  \at{The effect of
  secondary motion on axial transport in oscillatory tube flow}.  \jt{J. Fluid
  Mech.}  \bvol{193},  \pg{347--367}.

\bibitem[Poddar {\em et~al.\/}(2024)Poddar, Saha, Mondal, Dhar \&
  Mazumder]{poddar_effect_2024}
{\sc \au{Poddar, N.}, \au{Saha, G.}, \au{Mondal, K.~K.}, \au{Dhar, S.} \&
  \au{Mazumder, B.~S.}} \yr{2024}  \at{Effect of phase exchange kinetics on
  {{Taylor}} dispersion of chemically reactive solutes in an oscillatory
  magnetohydrodynamics flow between two parallel plates}.  \jt{Phys. Fluids}
  \bvol{36}~(5),  \pg{053601}.

\bibitem[Rana \& Murthy(2016)]{rana_solute_2016}
{\sc \au{Rana, J.} \& \au{Murthy, P. V. S.~N.}} \yr{2016}  \at{Solute
  dispersion in pulsatile {{Casson}} fluid flow in a tube with wall
  absorption}.  \jt{J. Fluid Mech.}  \bvol{793},  \pg{877--914}.

\bibitem[Shapiro \& Brenner(1987)]{shapiro_convection_1987}
{\sc \au{Shapiro, M.} \& \au{Brenner, H.}} \yr{1987}  \at{Convection and
  diffusion accompanied by bulk and surface chemical reactions in time-periodic
  one-dimensional flows}.  \jt{SIAM J. Appl. Math.}  \bvol{47}~(5),
  \pg{1061--1075}.

\bibitem[Shapiro \& Brenner(1990)]{shapiro_taylor_1990a}
{\sc \au{Shapiro, M.} \& \au{Brenner, H.}} \yr{1990}  \at{Taylor dispersion in
  the presence of time-periodic convection phenomena. {{Part I}}.
  {{Local}}-space periodicity}.  \jt{Phys. Fluids Fluid Dyn.}  \bvol{2}~(10),
  \pg{1731--1743}.

\bibitem[Singh \& Murthy(2023)]{singh_significance_2023}
{\sc \au{Singh, S.} \& \au{Murthy, P. V. S.~N.}} \yr{2023}  \at{Significance of
  skewness and kurtosis on the solute dispersion in pulsatile
  {{Carreau}}--{{Yasuda}} fluid flow in a tube with wall absorption}.  \jt{J.
  Fluid Mech.}  \bvol{962},  \pg{A42}.

\bibitem[Smith(1982{\natexlab{{\em a\/}}})]{smith_contaminant_1982}
{\sc \au{Smith, R.}} \yr{1982{\natexlab{{\em a\/}}}}  \at{Contaminant
  dispersion in oscillatory flows}.  \jt{J. Fluid Mech.}  \bvol{114},
  \pg{379--398}.

\bibitem[Smith(1982{\natexlab{{\em b\/}}})]{smith_dispersion_1982}
{\sc \au{Smith, R.}} \yr{1982{\natexlab{{\em b\/}}}}  \at{Dispersion of tracers
  in the deep ocean}.  \jt{J. Fluid Mech.}  \bvol{123},  \pg{131--142}.

\bibitem[Smith(1983)]{smith_contraction_1983}
{\sc \au{Smith, R.}} \yr{1983}  \at{The contraction of contaminant
  distributions in reversing flows}.  \jt{J. Fluid Mech.}  \bvol{129},
  \pg{137--151}.

\bibitem[Song \& Law(2015)]{song_longitudinal_2015}
{\sc \au{Song, J.} \& \au{Law, A. W.-K.}} \yr{2015}  \at{Longitudinal
  dispersion of turbulent oscillatory pipe flows}.  \jt{Environ. Fluid Mech.}
  \bvol{15}~(3),  \pg{563--593}.

\bibitem[Taylor(1953)]{taylor_dispersion_1953}
{\sc \au{Taylor, G.~I.}} \yr{1953}  \at{Dispersion of soluble matter in solvent
  flowing slowly through a tube}.  \jt{Proc. R. Soc. Lond. A}
  \bvol{219}~(1137),  \pg{186--203}.

\bibitem[Vedel \& Bruus(2012)]{vedel_transient_2012}
{\sc \au{Vedel, S.} \& \au{Bruus, H.}} \yr{2012}  \at{Transient
  {{Taylor}}--{{Aris}} dispersion for time-dependent flows in straight
  channels}.  \jt{J. Fluid Mech.}  \bvol{691},  \pg{95--122}.

\bibitem[Vedel {\em et~al.\/}(2014)Vedel, Hovad \&
  Bruus]{vedel_time-dependent_2014}
{\sc \au{Vedel, S.}, \au{Hovad, E.} \& \au{Bruus, H.}} \yr{2014}
  \at{Time-dependent {{Taylor}}--{{Aris}} dispersion of an initial point
  concentration}.  \jt{J. Fluid Mech.}  \bvol{752},  \pg{107--122}.

\bibitem[Vedel {\em et~al.\/}(2010)Vedel, Olesen \&
  Bruus]{vedel_pulsatile_2010}
{\sc \au{Vedel, S.}, \au{Olesen, L.~H.} \& \au{Bruus, H.}} \yr{2010}
  \at{Pulsatile microfluidics as an analytical tool for determining the dynamic
  characteristics of microfluidic systems}.  \jt{J. Micromechanics
  Microengineering}  \bvol{20}~(3),  \pg{035026}.

\bibitem[Wang {\em et~al.\/}(2021)Wang, Jiang, Chen, Tao \&
  Li]{wang_vertical_2021}
{\sc \au{Wang, B.}, \au{Jiang, W.}, \au{Chen, G.}, \au{Tao, L.} \& \au{Li, Z.}}
  \yr{2021}  \at{Vertical distribution and longitudinal dispersion of
  gyrotactic microorganisms in a horizontal plane {{Poiseuille}} flow}.
  \jt{Phys. Rev. Fluids}  \bvol{6}~(5),  \pg{054502}.

\bibitem[Wang \& Chen(2015)]{wang_environmental_2015}
{\sc \au{Wang, P.} \& \au{Chen, G.~Q.}} \yr{2015}  \at{Environmental dispersion
  in a tidal wetland with sorption by vegetation}.  \jt{Commun. Nonlinear Sci.
  Numer. Simul.}  \bvol{22}~(1),  \pg{348--366}.

\bibitem[Wang \& Chen(2017{\natexlab{{\em a\/}}})]{wang_basic_2017}
{\sc \au{Wang, P.} \& \au{Chen, G.~Q.}} \yr{2017{\natexlab{{\em a\/}}}}
  \at{Basic characteristics of {{Taylor}} dispersion in a laminar tube flow
  with wall absorption: exchange rate, advection velocity, dispersivity,
  skewness and kurtosis in their full time dependance}.  \jt{Int. J. Heat Mass
  Transf.}  \bvol{109},  \pg{844--852}.

\bibitem[Wang \& Chen(2017{\natexlab{{\em b\/}}})]{wang_concentration_2017}
{\sc \au{Wang, P.} \& \au{Chen, G.~Q.}} \yr{2017{\natexlab{{\em b\/}}}}
  \at{Concentration distribution for pollutant dispersion in a reversal laminar
  flow}.  \jt{J. Hydrol.}  \bvol{551},  \pg{151--161}.

\bibitem[Watson(1983)]{watson_diffusion_1983}
{\sc \au{Watson, E.~J.}} \yr{1983}  \at{Diffusion in oscillatory pipe flow}.
  \jt{J. Fluid Mech.}  \bvol{133},  \pg{233--244}.

\bibitem[Wu \& Chen(2014)]{wu_approach_2014}
{\sc \au{Wu, Z.} \& \au{Chen, G.~Q.}} \yr{2014}  \at{Approach to transverse
  uniformity of concentration distribution of a solute in a solvent flowing
  along a straight pipe}.  \jt{J. Fluid Mech.}  \bvol{740},  \pg{196--213}.

\bibitem[Wu {\em et~al.\/}(2012)Wu, Zeng, Chen, Li, Shao, Wang \&
  Jiang]{wu_environmental_2012}
{\sc \au{Wu, Z.}, \au{Zeng, L.}, \au{Chen, G.~Q.}, \au{Li, Z.}, \au{Shao, L.},
  \au{Wang, P.} \& \au{Jiang, Z.}} \yr{2012}  \at{Environmental dispersion in a
  tidal flow through a depth-dominated wetland}.  \jt{Commun. Nonlinear Sci.
  Numer. Simul.}  \bvol{17}~(12),  \pg{5007--5025}.

\bibitem[Yang {\em et~al.\/}(2021)Yang, Jiang, Wu, Wang, Wu, Zhang \&
  Zeng]{yang_migration_2021}
{\sc \au{Yang, Y.}, \au{Jiang, W.~Q.}, \au{Wu, Y.~H.}, \au{Wang, P.}, \au{Wu,
  Z.}, \au{Zhang, B.} \& \au{Zeng, L.}} \yr{2021}  \at{Migration of
  buoyancy-controlled active particles in a laminar open-channel flow}.
  \jt{Adv. Water Resour.}  \bvol{156},  \pg{104023}.

\bibitem[Yasuda(1982)]{yasuda_longitudinal_1982}
{\sc \au{Yasuda, H.}} \yr{1982}  \at{Longitudinal dispersion due to the
  boundary layer in an oscillatory current: theoretical analysis in the case of
  an instantaneous line source}.  \jt{J. Oceanogr. Soc. Jpn.}  \bvol{38}~(6),
  \pg{385--394}.

\bibitem[Yasuda(1984)]{yasuda_longitudinal_1984}
{\sc \au{Yasuda, H.}} \yr{1984}  \at{Longitudinal dispersion of matter due to
  the shear effect of steady and oscillatory currents}.  \jt{J. Fluid Mech.}
  \bvol{148},  \pg{383--403}.

\bibitem[Young {\em et~al.\/}(1982)Young, Rhines \&
  Garrett]{young_shear-flow_1982}
{\sc \au{Young, W.~R.}, \au{Rhines, P.~B.} \& \au{Garrett, C. J.~R.}} \yr{1982}
   \at{Shear-flow dispersion, internal waves and horizontal mixing in the
  ocean}.  \jt{J. Phys. Oceanogr.}  \bvol{12}~(6),  \pg{515--527}.

\bibitem[Zeng {\em et~al.\/}(2012)Zeng, Chen, Wu, Li, Wu \& Ji]{zeng_flow_2012}
{\sc \au{Zeng, L.}, \au{Chen, G.~Q.}, \au{Wu, Z.}, \au{Li, Z.}, \au{Wu, Y.~H.}
  \& \au{Ji, P.}} \yr{2012}  \at{Flow distribution and environmental
  dispersivity in a tidal wetland channel of rectangular cross-section}.
  \jt{Commun. Nonlinear Sci. Numer. Simul.}  \bvol{17}~(11),  \pg{4192--4209}.

\end{thebibliography}

\end{document}